\documentclass[twocolumn]{aastex62}
\usepackage{amsmath}
\revised{August 6, 2019}
\accepted{August 8, 2019}

\shorttitle{Moreton wave on 2014 March 29}
\shortauthors{Cabezas et al.}

\begin{document}

\title{Dynamic Processes of the Moreton Wave on 2014 March 29}

\correspondingauthor{Denis Pavel Cabezas}
\email{denis@kwasan.kyoto-u.ac.jp}

\author[0000-0002-0786-7307]{Denis P. Cabezas}
\affil{Astronomical Observatory, Kyoto University, Yamashina, Kyoto, 607-8471, Japan}
\affiliation{Department of Astronomy, Kyoto University, Sakyo, 606-8502, Japan}

\author{Ayumi Asai}
\affil{Astronomical Observatory, Kyoto University, Yamashina, Kyoto, 607-8471, Japan}

\author{Kiyoshi Ichimoto}
\affil{Astronomical Observatory, Kyoto University, Yamashina, Kyoto, 607-8471, Japan}

\author{Takahito Sakaue}
\affil{Astronomical Observatory, Kyoto University, Yamashina, Kyoto, 607-8471, Japan}
\affiliation{Department of Astronomy, Kyoto University, Sakyo, 606-8502, Japan}

\author{Satoru UeNo}
\affil{Astronomical Observatory, Kyoto University, Yamashina, Kyoto, 607-8471, Japan}

\author{Jos\'{e} K. Ishitsuka}
\affiliation{Universidad Nacional del Centro del Per\'{u}, Av. Ramon Castilla 3909, Huancayo, Per\'{u}}

\author{Kazunari Shibata}
\affil{Astronomical Observatory, Kyoto University, Yamashina, Kyoto, 607-8471, Japan}


\begin{abstract}
On 2014 March 29, an intense solar flare classified as X1.0 occurred in the active region 12017.
Several associated phenomena accompanied this event, among them a fast-filament eruption, large-scale
propagating disturbances in the corona and the chromosphere including a Moreton wave, and a coronal
mass ejection. This flare was successfully detected in multiwavelength imaging in H$\alpha$ line by
the Flare Monitoring Telescope (FMT) at Ica University, Peru.
We present a detailed study of the Moreton wave associated with the flare in question. Special attention
is paid to the Doppler characteristics inferred from the FMT wing (H$\alpha$ $\pm0.8$~{\AA}) observations,
which are used to examine the downward/upward motion of the plasma in the chromosphere.
Our findings reveal that the downward motion of the chromospheric material at the front of the Moreton wave
attains a maximum velocity of $4~\rm{km~s^{-1}}$, whereas the propagation speed ranges between $640$ and
$859~{\rm~km~s^{-1}}$.
Furthermore, utilizing the weak shock approximation in conjunction with the velocity amplitude of the
chromospheric motion induced by the Moreton wave, we derive the Mach number of the incident shock in the corona.
We also performed the temperature-emission measure analysis of the coronal wave based on the Atmospheric Imaging
Assembly (AIA) observations, which allowed us to derive the compression ratio, and to estimate the Alfv\'en
and fast-mode Mach numbers of the order of 1.06--1.28 and 1.05--1.27.
Considering these results and the MHD linear theory we discuss the characteristics of the shock front and the
interaction with the chromospheric plasma.
\end{abstract}



\section{Introduction} \label{sec:intro}
Shock waves associated with explosive events are fundamental physical processes in the solar atmosphere,
whose properties and effects have widely been discussed in the framework of magnetohydrodynamics (MHD) theory.
In the Sun's chromosphere a large-scale wavelike propagating disturbance, that was discovered by
\citet{moreton1960} and \citet{mramsey1960} during the observation of a solar flare in the wing of the H$\alpha$
line, occasionally happens in association with strong flares. 
This wave phenomenon known as ``Moreton wave" usually propagates as semicircular fronts through the chromosphere
over a distance of about $5\times10^{5}~\rm{km}$ from the flaring region, at the speed of the order of
$500 - 1500~\rm{km~s}^{-1}$. Associated with Moreton waves, filaments and prominences standing in the way of
propagation direction can abruptly be induced to oscillate, and in some situations to a disruption. Observations
in H$\alpha$ spectral line reveal that the leading front of the Moreton wave is observed as enhanced absorption
in the red-wing and as reduced absorption (brightening) in the blue-wing, followed by a fainter reversed front,
as enhanced absorption in the blue-wing and as reduced absorption in the red-wing \citep{dodson1968}. In the core
of H$\alpha$ line the wave fronts usually appear as brightening.
These observational characteristics were interpreted as depression and a subsequent relaxation process of the
chromospheric material, as a consequence of the plasma being forced to move downward with a velocity of
$6 - 10~\rm{km~s}^{-1}$~\citep{svestka1976} inferred from the motions of spicules and fibrils caused by the wave
front passage.

The kinematical properties of the Moreton wave such as the high speed, the great traveling distances, and the
low wave amplitude, led to the question whether or not this phenomenon is of chromospheric origin; since the
mentioned physical parameters are incompatible with the framework of wave propagation in the chromosphere.
For example, for a typical value of the temperature in the chromosphere of $10^{4}~\rm{K}$, the sound speed results
in $\sim$$10~\rm{km~s^{-1}}$, and assuming a density $10^{11}~\rm{cm^{-3}}$ and $15~\rm{G}$ of magnetic field,
the Alfv\'{e}n speed in the chromosphere is about $100~\rm{km~s^{-1}}$. A wave traveling in the chromosphere
with a speed of $1000~\rm{km~s^{-1}}$, as determined from observations, would imply that this wave moves
with a Mach number of the order of 10. A shock wave of such strength should exhibit a strong amplitude and would
fully be dissipated before traveling long distances, however, those effects are not observed.
A comprehensive interpretation on the generation and propagation aspects of the Moreton wave is conducted if wave
propagation in the corona is considered, where the sound and Alfv\'{e}n speeds are about one order of magnitude
larger than in the chromosphere. This idea was initially postulated by \citet{anderson1966} and \citet{meyer1968},
and further modeled by \citet{uchida1968}. According to Uchida's model, the pressure pulse during the explosive
phase of a solar flare generates an MHD fast-mode wave in the corona, which compresses and pushes downward the
chromospheric plasma, and as a result the Moreton wave is created. A good correlation between type II radio burst
and Moreton waves does exit \citep{kai1970}. Since the former traces the upward propagating fast-mode shock in the
corona, it supports the Uchida's scenario.

In 1997, a large-scale wave propagating disturbance was discovered in coronal emission line Fe~{\sc xii} (195~\AA)
in association with an C-class flare~\citep{thompson1999}. This coronal wave, named as ``EIT wave" after the
Extreme-ultraviolet Imaging Telescope \cite[EIT;][]{delabou1995} on board the
{\it Solar and Heliospheric Observatory}~\cite[{\it SOHO};][]{domingo1995}, was expected to be the fast-mode MHD
counterpart of the Moreton wave as predicted by \citet{uchida1968}. However, over the last 20 years there was an
intense debate in interpreting this phenomenon as an MHD fast-mode wave \cite[see reviews by][]{chen2016,long2017}.
The controversy resides in its physical nature, such as the long-lasting coronal dimming, stationary brightening,
and particularly its traveling speed; since it ranges between
$170-350~\rm{km~s^{-1}}$~\cite[][]{klassen2000,thompson2009} being slower than the expected fast-mode speed, and
sometimes even slower than the sound speed in the corona \citep{thompson2009}. In addition, the EIT waves lack the
correlation with type II radio bursts, whose speed is in the range of $300-1200~\rm{km~s^{-1}}$~\citep{klassen2000}.
In order to overcome the discrepancies on the nature of the EIT and Moreton waves, several models and possibilities
were put forward. For example, based on MHD simulations by \citet{chen2002} and extending them, \citet{chen2005a}
found that an upward-expanding flux-rope is capable to trigger a coronal mass ejection (CME) by which
two wavelike propagating disturbances are produced: a fast-mode shock wave, and a slower component related to the
density perturbation resulting from successive stretching field lines that overlies the erupting flux-rope. 
Here it is worth noticing that \cite{delannee1999} identified a bright front in EIT images associated with a
flare and CME, and they proposed a scenario of magnetic field reconfiguration as a model of EIT waves.
Alternative interpretations have also been suggested in order to clarify the inconsistency of the velocity between
the EIT and Moreton waves, such as a subsequent deceleration \citep{warmuth2001}, which was explained in terms of
fast-mode shock produced by a ``blast wave" rather than a flux-rope expansion.
Note that in the meantime, large-scale disturbances associated with flares traveling through the corona were also
detected in soft X-ray emission \cite[e.g.,][]{khan-aurass2002,narukage2002,hudson2003,warmuth2005,asai2008}, 
and reported in radio wavelengths as type~II burst \cite[e.g.,][]{khan-aurass2002,eto2002,vrsnak2005}. They were
interpreted as the coronal counterpart of Moreton waves.

Nearly a decade, the Atmospheric Imaging Assembly \cite[AIA;][]{lemen2012} instrument on board the
{\it Solar Dynamics Observatory} \cite[{\it SDO};][]{pesnell2012} is constraining our understanding of coronal waves.
From now on, in this paper we will use the term ``EUV waves" for referring large-scale coronal disturbances observed
in extreme ultraviolet (EUV) passbands. Thanks to the AIA's unprecedented capability, significant discoveries were
possible, including intrinsic characteristics never seen before of the coronal EUV waves. So, it is more likely that
the controversy on the EIT waves is coming to an end.
Nowadays, it is being more accepted that two kinds of large-scale wavelike phenomena appear in the corona in
association with flares and CMEs. The first, the MHD fast-mode or shock traveling ahead, and the second or slow
component straddling behind. This picture was initially reported by \citet{harra2003} based on TRACE \citep{handy1999}
observations, suggested by analytical and numerical approaches \cite[e.g.,][]{chen2002}, and is being confirmed
by numerous recent studies \cite[cf.][]{liu2014,warmuth2015}. It was also found a clear evidence of co-spatiality
between the fast-mode EUV wave and the Moreton wave~\citep{asai2012}, and the association with type~II radio burst
displays a high tendency \citep{nitta2013}. It should be mentioned that statistical investigations have shown that
EUV waves are more linked to CMEs rather than solar flares~\citep{biesecker2002,nitta2013,nitta2014,muhr2014},
although the mechanism by which they are related still remains puzzling.

The 2014 March 29 flare is one of the best observed event, because it was captured in a wide range of wavelengths
with unprecedented temporal and spatial resolution by advanced space-borne and ground-based instruments. This powerful
flare classified as X1.0 on the {\it GOES} scale occurred in the active region 12017 (N11, W32), and was accompanied
by various associated phenomena: a fast-filament eruption, chromospheric evaporation, large-scale propagating
disturbance in the corona and the chromosphere, filaments activation/oscillation, and ultimately a CME.
A number of studies have been conducted focusing on different aspects of this flare. For example, \citet{kleint2015}
investigated the rapid filament eruption and its role in the flare generation, \citet{battaglia2015} have concentrated
in the electron beams as the driver of the chromospheric evaporation, \citet{aschwanden2015} estimated the magnetic
energy dissipation, while \citet{francile2016} reported on the kinematics of the Moreton and EUV waves in association
with the CME. Here in this paper we concentrate on the chromospheric responses to the coronal wave associated with
the flare in question. Particular attention is paid to the Doppler characteristics of the Moreton wave inferred from
H$\alpha$ wing observations, which are used to examine the downward/upward motions of the chromospheric plasma and its
connection with the propagating disturbance in the corona.

In an effort to better understand the nature of the Moreton waves, several but a few studies have been conducted
using H$\alpha$~wing observations \cite[e.g.,][]{warmuth2004,veronig2006,balasubramaniam2007,narukage2008,muhr2010,
balasubramaniam2010}, also in H$\beta$ \citep{zhang2001}, and in Helium~I line at 10830~{\AA}
\cite[][]{vrsnak2002a,gilbert2004}. Most of the studies have focused on the kinematics of the horizontal and lateral
propagation speed, and some of them on the morphology and Doppler velocity of the wave fronts, although, no firm
conclusions were drawn about the latter. For example, \cite{balasubramaniam2007} based on Dopplergrams of a Moreton
wave reported a Doppler velocity of about $2.6~{\rm km~s^{-1}}$ in the chromosphere.
Certainly, the signature of Moreton waves is very faint, and in general the perturbation or the associated
line-shift of the H$\alpha$ spectral profile cannot easily be distinguished.

\section{Observations and Data}
The Moreton wave on 2014 March 29, was detected in high cadence images obtained by the Flare Monitoring Telescope
\citep[FMT:][]{kurokawa1995,ueno2007,shibata2011}, in operation at National University San Luis Gonzaga, Ica, Peru.
The FMT provides simultaneously full-disk solar images at several wavelengths: H$\alpha$ line center (6562.8~{\AA}),
H$\alpha$ $-0.8$~(6562.0~{\AA}; blue-wing) and H$\alpha$ $+0.8$ (6563.6~{\AA}; red-wing), and continuum (6100~{\AA}),
also in prominence mode (H$\alpha$ line center with occulting disk), with a time cadence of $20~\rm{s}$ and spatial 
sampling of $\sim2\farcs0~\rm{pixel}^{-1}$. These characteristics make FMT a suitable instrument to detect Moreton 
waves \citep[e.g.,][]{eto2002,narukage2002,narukage2004}, to derive the velocity field of filaments eruption 
\citep[e.g.,][]{morimoto2003a,morimoto2003b, morimoto2010,cabezas2017}, and to study the long-term variation of UV 
radiation (UeNo et al. 2019, in prep). Since Moreton waves are best observed in the wing of the H$\alpha$ line 
\citep[cf.][]{mramsey1960,narukage2008,chen2005b}, for the analysis we made use of FMT wing (red \& blue) observations.
This further enable us to investigate the Doppler characteristics of the Moreton wave.

\begin{figure}
\centering
\includegraphics[scale=0.85]{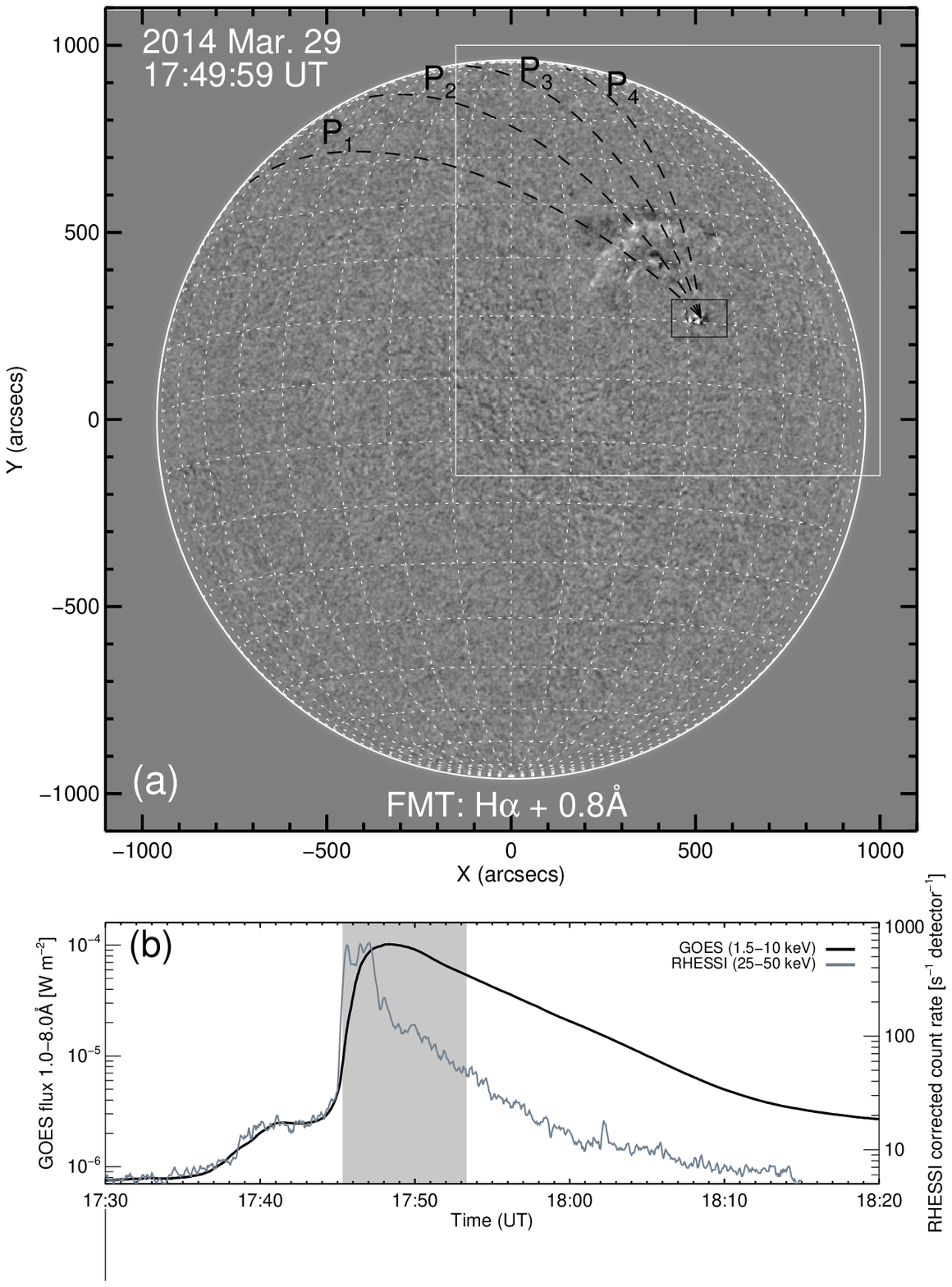}
	\caption{ {\small (a) Time-difference image of full-disk Sun at H$\alpha$ $+0.8$~{\AA} captured by the
	Flare Monitoring Telescope (FMT) on 2014 March 29 at 17:49:59~UT. The image was subtracted with its
	precedent taken at 17:49:20~UT. The dashed black lines labeled P$_{1}$, P$_{2}$, P$_{3}$, and P$_{4}$
	projected on the solar surface, outlines the paths along which the analysis of the Moreton wave is performed.
	The black rectangle marks the flare site, while the large white box depicts the field of view presented
	in Figure~\ref{fig:FIG2}. (b) GOES X-ray emission at 1.0--8.0~\AA~(1.5--10~keV, black solid line) and
	RHESSI count rate (25--50~keV, gray solid line) of the X1.0 flare. The shaded area starting from 17:45:20
	to 17:53:20~UT, denotes the time period at which the wave front is identified in our H$\alpha$ data. 
	The animation shows time series running-difference maps of the Moreton wave at $+0.8$ and $-0.8$~{\AA},
	from 17:42 to 18:00~UT.}
        \label{fig:FIG1}}
\end{figure}

In Figure~\ref{fig:FIG1}a we present a full-disk time-difference image at H$\alpha$ $+0.8$~{\AA} (red-wing), taken
by the FMT on 2014 March 29 at 17:49:20 and 17:49:59~UT. The Moreton wave is clearly seen as an arc-shaped
front propagating north-east from the flare site following a trajectory of a solid angle. The great-circle dashed
lines projected on the solar surface outline the paths P$_{1}$, P$_{2}$, P$_{3}$, and P$_{4}$ considered in our
analysis to estimate the horizontal propagation speed of the Moreton wave and to derive the velocity amplitude in
the chromosphere. Figure~\ref{fig:FIG1}b shows the soft X-ray flux by {\it GOES} together with the hard X-ray
emission by {\it RHESSI} in the energy range 25--50~keV. According to the {\it GOES} plot, which is predominantly
thermal emission from the corona, the flare started at $\sim$17:35:30~UT reaching its maximum at $\sim$17:48:00~UT;
whereas the {\it RHESSI} profile shows a different behavior during the impulsive and the main phase of the flare
because it results from non-thermal bremsstrahlung emission. Figure~\ref{fig:FIG1}b also indicates the time period
at which the Moreton wave is observed, as represented by the gray shaded area extending from 17:45:20 to 17:53:20~UT.

The coronal wave associated with the flare under study was captured in X-ray wavelength by the X-Ray Telescope
\citep[XRT;][]{golub2007} on board the {\it Hinode} satellite~\citep{kosugi2007}. We use full frame images through
the Ti-poly filter with a temporal resolution of $\sim$$40~\rm{s}$ and pixel size $\sim$$8\farcs2~\rm{pixel}^{-1}$,
and partial images in Al-poly and  Be-thin filters with a temporal resolution and spatial sampling of $\sim$$1~\rm{min}$
and $\sim$$1\farcs0~\rm{pixel}^{-1}$. The XRT channels are sensitive to very hot coronal plasma capable to detect
emissions from $\sim$$1$ to more than $10~\rm{MK}$.
For our study we also used extreme ultraviolet (EUV) data taken by the Atmospheric Imaging Assembly 
\citep[AIA;][]{lemen2012} on board the {\it Solar Dynamics Observatory} \citep[{\it SDO};][]{pesnell2012}. 
We examine the morphology and the kinematics of the coronal wave by using emission lines at 304, 211, and 94~\AA, 
which are dominated by He~{\sc ii}~($\log~T$$\sim$$4.7$), Fe~{\sc xiv}~($\log~T$$\sim$$6.3$), and Fe~{\sc xviii}
($\log~T$$\sim$$6.8$), respectively. To characterize the local plasma responses to the EUV wave passage, we carry out
differential emission measure (DEM) analysis based on the method introduced by \citet{cheung2015}. We performed DEM
maps using six AIA (94, 131, 171, 193, 211, 335~\AA) coronal temperature data, with full cadence~($12~\rm{s}$) and
spatial sampling of $\sim0\farcs6~\rm{pixel}^{-1}$.

\section{Wave Morphology and Kinematics}
The morphology of the Moreton wave and its associated coronal counterpart is characterized by performing
running-difference maps, while the kinematics is inferred from intensity variation through artificial slits and by
carrying out time-distance measurements of the wave front. For the kinematics we take into account great-circle paths
P$_{1}$--P$_{4}$ (shown in Figures~\ref{fig:FIG1}, \ref{fig:FIG2}), projected over an angular extent on the solar
surface in the wave propagation direction.

\subsection{Morphology}
\begin{figure}
\centering
\includegraphics[scale=0.75]{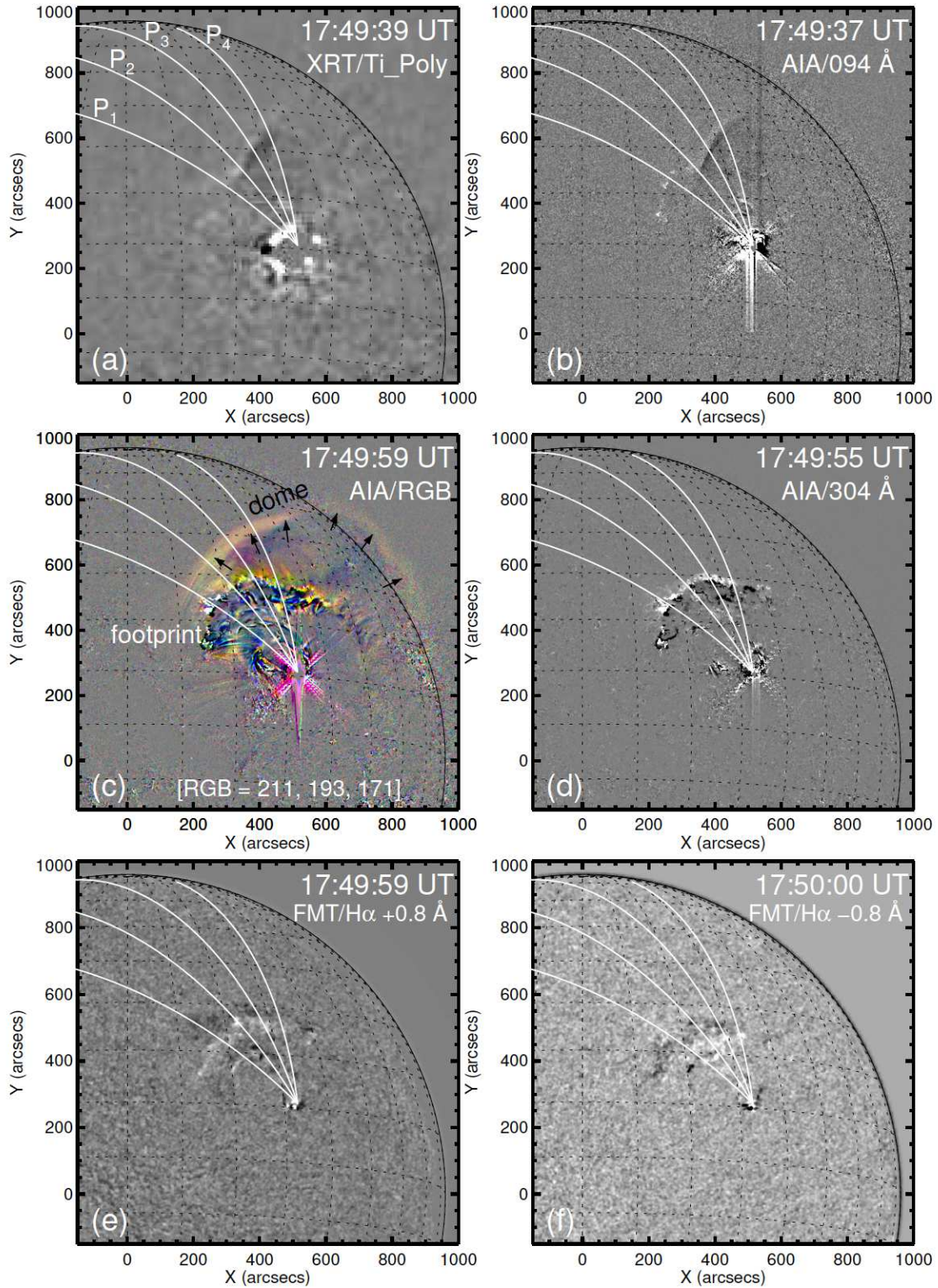}
	\caption{ {\small Coronal counterpart and Moreton wave associated with the X-class flare on 2014 March 29.
	The maps show different heights of the solar atmosphere, including the hotter component ($\sim$$6~{\rm MK}$)
	of corona in X-ray and 94~{\AA} (panels a, b), intermediate component ($1$--$2~{\rm MK}$) as a composition
	of AIA 211~{\AA}~(red), 193~{\AA}~(green) and 171~{\AA}~(blue) (panel c), transition region
	($\sim$$5\times10^{4}~{\rm K}$) at 304~{\AA} (panel d), and the chromosphere ($\sim$$10^{4}~{\rm K}$) at
	H$\alpha$~$\pm0.8$~{\AA} (panels e, f). The maps illustrate the response of the solar atmosphere to the global
	disturbance triggered by the large-scale wave propagation.
	In panel (c) the arrows highlight the dome-like expansion of the coronal wave propagating ahead of the
	disturbed surface (footprint). The animation includes full-disk frames of panel (c) and runs from 17:30
	to 17:59~UT.}}
        \label{fig:FIG2}
\end{figure}

Figure~\ref{fig:FIG2}a-c show time-difference coronal counterpart images of the Moreton wave on 2014~March~29
in X-ray channel (Ti-poly), in EUV band at 94~{\AA}, and a composite tri-color image (RGB = 211, 193, 171~\AA),
respectively. For comparison we also include time-difference images in He~{\sc ii} (304~{\AA})(panel~d) and
H$\alpha$ $\pm0.8$~{\AA} (panels e, f) to see how the different layers of the solar atmosphere respond to the
effects of the wave propagation. As the temperature changes from hot to cool, the disturbance
develops differently exhibiting several aspects and components of the coronal wave.

In X-ray and 94~{\AA} (Figure~\ref{fig:FIG2}a, b) the morphology of the coronal wave is quite similar in
space and time, both representing emissions of hotter coronal plasma $>$$6.0~\rm{MK}$. In Figure~\ref{fig:FIG2}c,
which is a composite map of warm (211, 193~{\AA}) and cool (171~{\AA}) AIA channels, it can be seen a dome-shaped
structure similar to a shock propagating ahead of the disturbed surface or its footprint, the latter
predominantly seen in the cool channel.
In much cooler line, i.e., He~{\sc ii} (Figure~\ref{fig:FIG2}d) which provides diagnostics of the upper chromosphere
and the transition region ($\sim$$5\times10^{4}~\rm{K}$), only the footprint of the coronal wave is clearly observed
as a bright arc-like feature moving towards the north-east. This thin layer possibly is strongly affected by the
pressure excess created by the globally propagation shock in the corona, resulting in such an enhanced emission.
On the other hand, the signature of the disturbance in the wing of the H$\alpha$ line at $\pm0.8$~{\AA}
(Figure~\ref{fig:FIG2}e, f) appears more diffuse and somewhat similar in morphology to that in He~{\sc ii} line.
Naturally, the  change observed in H$\alpha$ indicates an increase of the local Doppler velocity, and this aspect
will be addressed in section 4.

\subsection{Kinematics}
\begin{figure}
\centering
\includegraphics[scale=0.72]{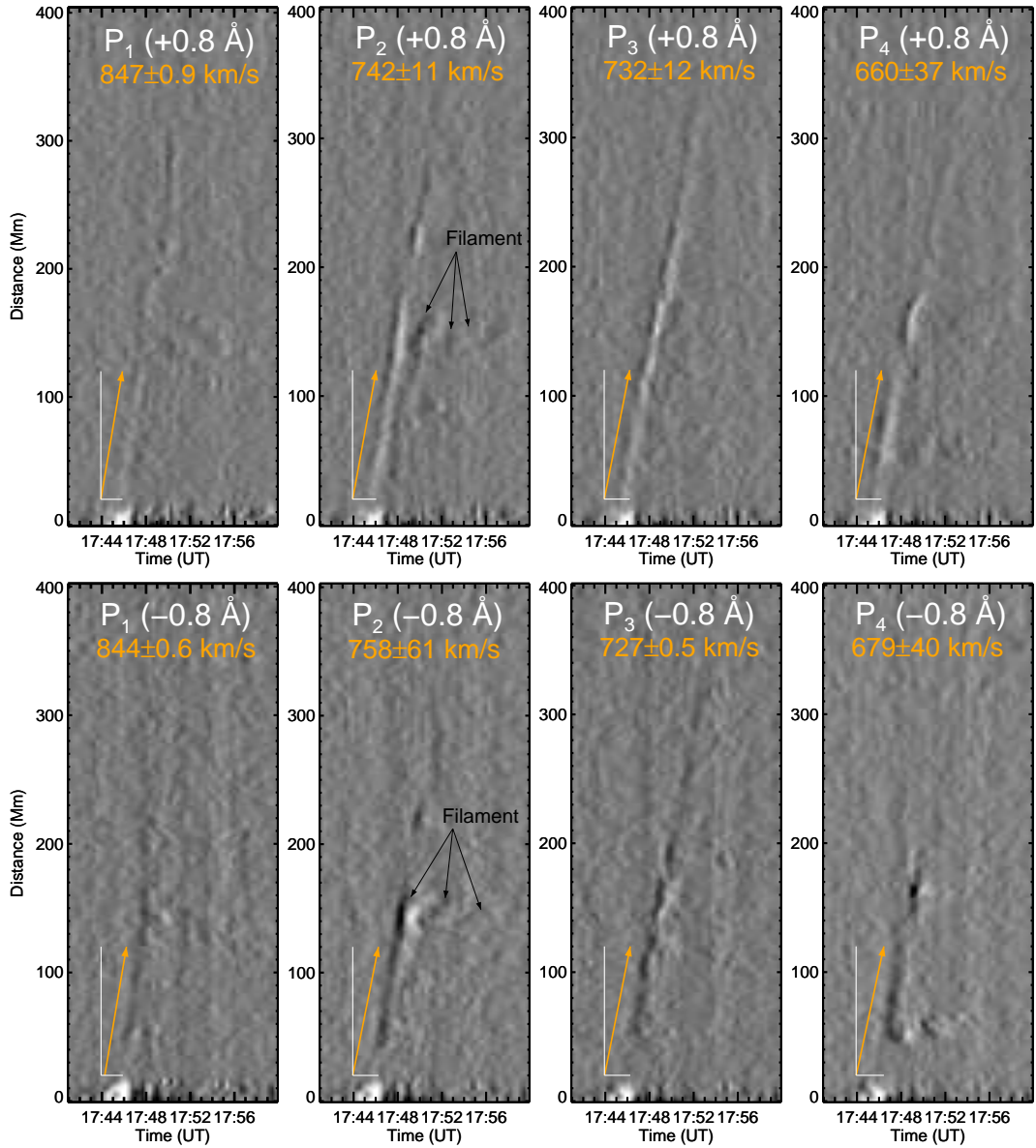}
	\caption{ {\small Time-distance diagrams of the Moreton wave obtained from running-difference images along
       trajectories P$_{1}$--P$_{4}$ (see panels e, f in Figure~\ref{fig:FIG2}). The distance is expressed in
	megameters (Mm) starting from the flare site. The upper panels show results at H$\alpha$ $+0.8$~{\AA}
        and the lower at H$\alpha$ $-0.8$~{\AA}, wherein the orange arrows in each panel outline the earlier propagation
	speed of the Moreton wave. On the other hand, the black arrows in panels P$_2$ point-out signatures of a filament
	activation due to the interaction with the Moreton wave}.} 
\label{fig:FIG3}
\end{figure}

We performed H$\alpha$ $-0.8$ and $+0.8$~{\AA} running-difference intensity maps in order to enhance the wave contrast.
Then, we extracted the intensity variation along the slits P$_1$--P$_4$ shown in Figure~\ref{fig:FIG2} (panels e, f).
The results are presented in Figure~\ref{fig:FIG3} as time-distance diagrams, in which the wave signature at
H$\alpha$ $+0.8$~{\AA} (upper panels) appears first as dark then as bright narrow stripes, while at H$\alpha$ $-0.8$~{\AA}
(lower panels) as bright and dark lanes (see also the animation of Figure~\ref{fig:FIG1}); both displaying
identical evolution and similar tendency in the propagation speed. In the time-distance diagram P$_{2}$, we can also
identify signatures of the wave front interacting with quiescent filaments located in the propagation direction of the
Moreton wave.

\begin{figure}
\centering
\includegraphics[scale=0.92]{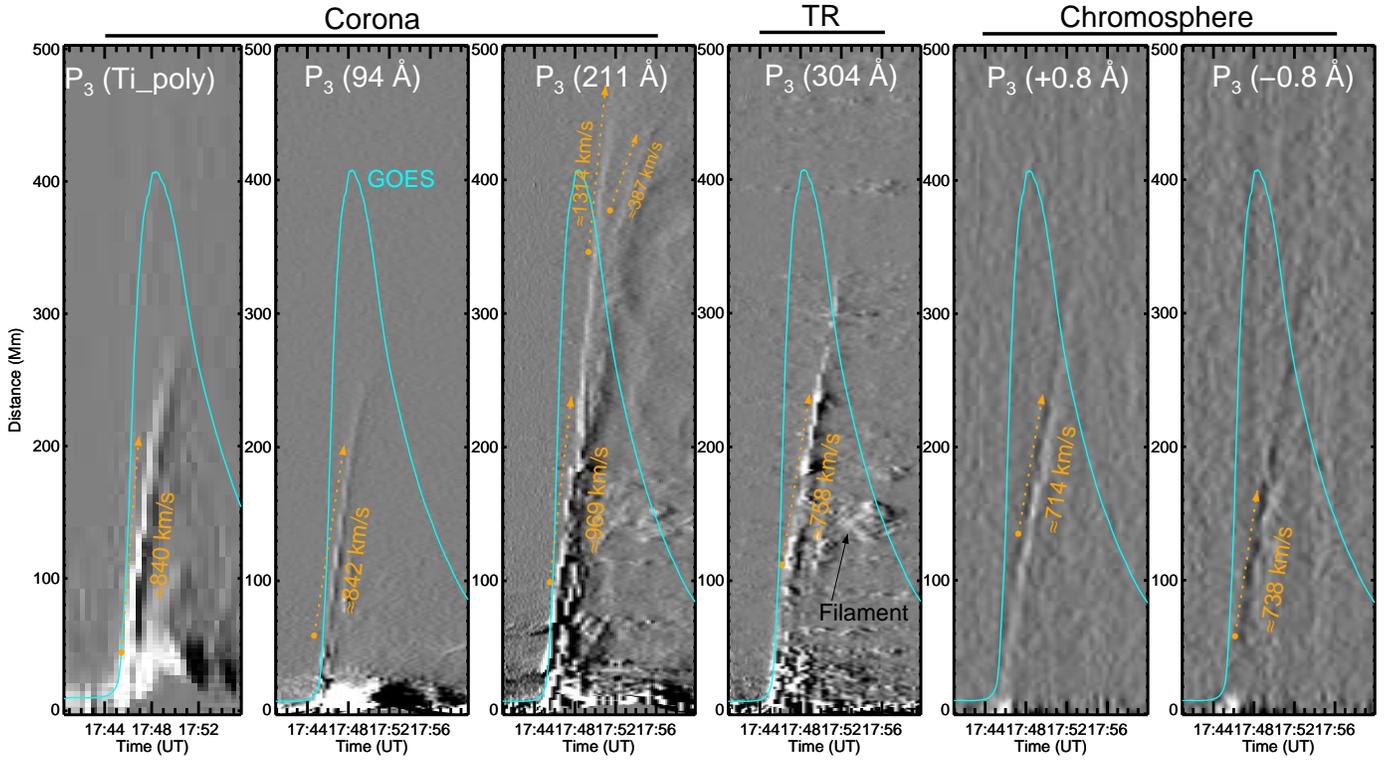}
	\caption{ {\small Time-distance diagrams of the disturbance in the corona, transition region (TR), and the
	chromosphere caused by the coronal wave on 2014 March 29. The intensity variation is calculated from
	running-difference images along trajectory P$_3$ shown in Figure~\ref{fig:FIG2}. From left to right: X-ray
        Ti-poly filter ($\log~T$$\sim$$6.7$, with short time-range), EUV at 94 ($\log~T$$\sim$$6.8$), 211 ($\log~T$$\sim$$6.3$),
	and 304~{\AA} ($\log~T$$\sim$$4.7$), and the wing of H$\alpha$ at $+0.8$, $-0.8$~{\AA}
        ($\log~T$$\sim$$4.0$), respectively. The cyan curves are the GOES X-ray emission at the 1--8~{\AA} channel.
        The time-distance diagrams reveal that the corona, transition region, and the chromosphere reacted almost
	simultaneously to the global scale wave propagation (see the text).} }
\label{fig:FIG4}
\end{figure}

In order to compare the nature of the Moreton wave with its associated coronal wave, time-distance diagrams from
EUV and X-ray observations are also performed, whose results obtained through slit P$_3$ (see Figure~\ref{fig:FIG2})
are shown in Figure~\ref{fig:FIG4}. The following characteristics can be drawn from the multiwavelength
time-distance diagrams:

\begin{itemize}
	\item{Taking the starting time of the GOES flux enhancement as reference, the corona, transition region,
                and the chromosphere reacted almost simultaneously to the extremely large propagating disturbance
                in the solar atmosphere, even though we note some delay in the response of more dense layers, i.e.,
                the transition region and the chromosphere.}

	\item{A similarity of the wave signature in terms of velocity and development is observed in H$\alpha$
		and He~{\sc ii}~(304~{\AA}) lines, although in the latter the signature appears brighter and
		sharper (see also panel d in Figure~\ref{fig:FIG2}). The enhanced emission in He~{\sc ii} suggests
		that the transition region is evidently more affected than the chromosphere by the pressure jump as
		a consequence of the coronal wave passage.} 

        \item{Several characteristics at 211~{\AA} are identified. The main or fast-component traveling at
                $\sim$$969\rm~{km~s^{-1}}$ and a slow-component moving behind it with a speed of about
                $387\rm~{km~s^{-1}}$. Here it is worth to add some comments on the observed features. The rapid
                expansion and the large propagation speed is an inherent characteristic of a fast-mode magnetosonic
                wave, in contrast the slower component possibly results from the deformation and restructuring of
                the ambient corona after the shock passage.}

	\item{In the AIA hotter channel (94~{\AA}) and X-ray Ti-poly the wave progression behaves identically,
		although the intensity variation due to the disturbance lasts only for about $6$~minutes.
                The observed short period in these hotter channels can be ascribed to the temperature response
                of the instruments, since they are sensitive to hotter coronal emission well above $6~{\rm MK}$.
                Therefore, it is more likely that the 94~{\AA} and Ti-poly filters detected mainly the hottest
                part of the coronal wave, possibly present during the early stage of the wave propagation.}
\end{itemize}

\begin{figure}
\centering
\includegraphics[scale=0.82]{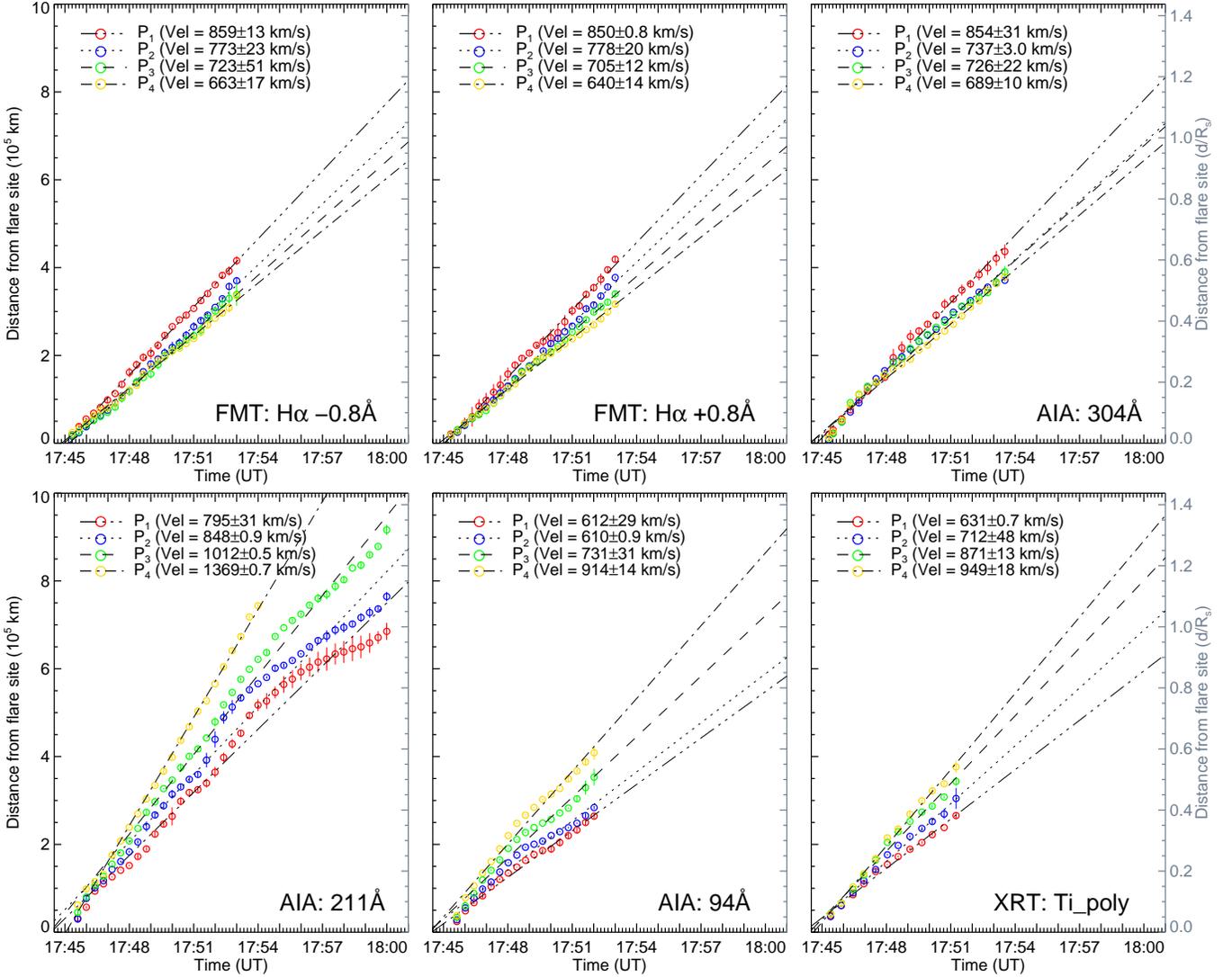}
	\caption{ {\small Time-distance plots of the Moreton wave and its associated coronal wave, measured from
	the flare site along paths P$_{1}$--P$_{4}$ shown in Figure~\ref{fig:FIG2}. The top panels show H$\alpha$ $-0.8$~{\AA}
	(blue-wing), H$\alpha$ $+0.8$~{\AA} (red-wing) and He~{\sc ii} (304~{\AA}), while the bottom panels show AIA 211 and
	94~{\AA}, and XRT Ti-poly filter. 
        It is also shown the corresponding mean velocity of the wave front derived from linear-fitting, the uncertainty
	is estimated from the standard deviation. The right vertical axis in each panel highlighted with gray color are in
	units of solar radii ($1{\rm R_{s}}\approx695,800~{\rm km}$).}} 
\label{fig:FIG5}
\end{figure}

On the other hand, we also estimated the propagating distances of the Moreton wave by tracking the leading edge
of the wave front in successive images. The measurements are done along four great-circle paths (P$_{1}$--P$_{4}$),
same trajectories for the time-distance diagrams, and assuming the flare site as the wave origin.
The mean propagation velocity of the Moreton wave is calculated by applying the linear fitting  to the time-distance
plots.
The results for H$\alpha$ $-0.8$ and $+0.8$~{\AA} are presented in Figure~\ref{fig:FIG5} (top panels), along
with the results of the wave progression derived from He~{\sc ii} line (304~{\AA}).
From the plots it can be seen that the Moreton wave moves faster along P$_{1}$ direction, both at H$\alpha$
$-0.8$ and $+0.8$~{\AA} with a mean velocity of $859$  and $850~\rm{km~s}^{-1}$, respectively.
It is observed that there is a tendency in the speed of the Moreton wave to slow-down as path changes from
the trajectory P$_{1}$ to P$_{4}$. The derived mean velocities along paths P$_{2}$, P$_{3}$, and P$_{4}$
vary from $773$ to $663\rm~{km~s}^{-1}$ at H$\alpha$ $-0.8$~{\AA}, and from $778$ to $640~\rm{km~s}^{-1}$
at H$\alpha$ $+0.8$~{\AA}, respectively.
The characteristics of the wave evolution at 304~{\AA} are somewhat identical to that in H$\alpha$. Such
identical features are clearly recognized by looking at the wave front speeds along paths P$_{1}$--P$_{4}$,
which are comparable to the derived speeds from H$\alpha$ observations, showing also a tendency of reduction
in the propagation speed as the wave moves apart from trajectory P$_{1}$ towards the northern-west.
The similarities of the wave progression in He~{\sc ii} and H$\alpha$ lines, and also in morphology as
described in section~3.1, is a clear indication that the plasma in the transition region and the chromosphere
reacted in similar manner to the action of the large-amplitude disturbance in the corona.

Similar analysis is conducted using AIA coronal lines at 211 and 94~{\AA}, as well as hotter coronal emission
captured by XRT Ti-poly filter. Figure~\ref{fig:FIG5}~(bottom panels) shows the time-distance plots derived
from 211~{\AA}, 94~{\AA}, and Ti-poly observations, measured along paths P$_{1}$--P$_{4}$, same as for H$\alpha$
and He~{\sc ii} lines.
It is shown that the wave in the corona develops in different manner in terms of speed and propagation direction.
For example, at 211~{\AA} the estimated values of the mean velocity extends from $795$ to $1369\rm~{km~s}^{-1}$,
which increases as the wave travels in the northern-west direction (paths P$_{3}$, P$_{4}$). Same tendency shows
the wave evolution at 94~{\AA} and in Ti-poly filter, having a mean velocity from $612$ to $914\rm~{km~s}^{-1}$,
and $631$ to $949\rm~{km~s}^{-1}$, respectively, although only the earlier stage of the wave front progression
is identified in these hotter channels.
Here it is worthwhile to point-out the following: because of the complex 3D dome-like expansion of the coronal
wave and the projection effect, the results presented in the bottom panels of Figure~\ref{fig:FIG5} do represent
the kinematics of the wave evolution at high altitudes rather than at the coronal base or the solar surface.
In Table~1 a summary of the propagation speeds of the Moreton wave and the associated coronal wave calculated
along the four paths are presented, including also the mean value for each wavelength domain.

\begin{deluxetable*}{ccccccc}[ht]
	\tablecaption{Propagation speeds of the Moreton wave and its associated coronal wave on 2014 March 29.
	\label{tab:mathmode}}
\tablecolumns{6}
\tablenum{1}
\tablewidth{0pt}
\tablehead{
\colhead{Path} & \colhead{H$\alpha$ $-0.8$~{\AA}} & \colhead{H$\alpha$ $+0.8$~{\AA}} & 
\colhead{304~{\AA}} & \colhead{211~{\AA}} & \colhead{94~{\AA}} & \colhead{XRT/Ti-poly}
}
\startdata
P$_{1}$ & 859~$\pm$~13        & 850~$\pm$~0.8 &854~$\pm$~30 & 795~$\pm$~31 & 612~$\pm$~29 & 631~$\pm$~0.7 \\
P$_{2}$ & 773~$\pm$~23        & 778~$\pm$~20  &737~$\pm$~3.0 & 848~$\pm$~0.9 & 610~$\pm$~0.9 & 712~$\pm$~48 \\
P$_{3}$ & 723~$\pm$~51        & 705~$\pm$~12  &726~$\pm$~22 &1012~$\pm$~0.5 & 731~$\pm$~31 & 871~$\pm$~13 \\
P$_{4}$ & 663~$\pm$~17        & 640~$\pm$~14  &689~$\pm$~10 &1369~$\pm$~0.7 & 914~$\pm$~14 & 949~$\pm$~18 \\
\hline
Mean   & 754.50               & 743.25        &751.50       &1006.00       & 716.75       & 790.75\\
\enddata
	\tablecomments{In units of ${\rm km~s^{-1}}$}
\end{deluxetable*}

\section{Doppler Characteristics of the Moreton Wave}
The Moreton wave is interpreted as a down-up swing disturbance in the chromosphere, as a result of the plasma
being pushed downward by the globally expanding fast-mode wave or shock in the corona.
This characteristic is regarded as compression followed by a relaxation process of the chromospheric material
and induces the Doppler shift of H$\alpha$ line and brightness change in off-band images. In this section we
investigate the Doppler characteristics of the Moreton wave.

\subsection{H$\alpha$ intensity profiles}
As a general picture, the signature of the Moreton wave in H$\alpha$ line appears as a dark front in absorption
in the red-wing and as bright front in the blue-wing, followed by a wider reversed disturbance in brightening and
darkening in the red and blue-wing, respectively. Since the FMT provides H$\alpha$ wing observations at two
wavelength of equal distance from the line center, the absorption and brightening patterns of the wave front can
be characterized by means of Doppler shift of the intensity profiles. To do so, we first normalized the
H$\alpha$~$+0.8$~($I_{red}$) and $-0.8$~{\AA}~($I_{blue}$) intensity maps to the spatially averaged quiet-Sun level
measured at disk center.
Next, considering the flare site as the wave origin and the four individual trajectories (P$_1$--P$_4$), same as
for the kinematics analysis described in section~3, intensity profiles as a function of distance $x$ and time $t$
are computed in red $I_{r}(x,t)$ and blue $I_{b}(x,t)$ wings by averaging the pixels intensities over consecutive
areas of $\sim$$44$~arcsecs$^{2}$~($1\times11$~pixels) along each trajectory. Because the interest here is to
determine the relative change of the intensity with respect to the unperturbed background, pre-event intensity
profiles $I_{r}(x,t_0)$, $I_{b}(x,t_0)$ at 17:42:39~UT are obtained at the same positions of each corresponding
trajectory, and subtracted from the intensity profiles; that is, $\Delta I_{r,b} = I_{r,b}(x,t) - I_{r,b}(x,t_0)$.

\begin{figure}
\centering
\includegraphics[scale=0.86]{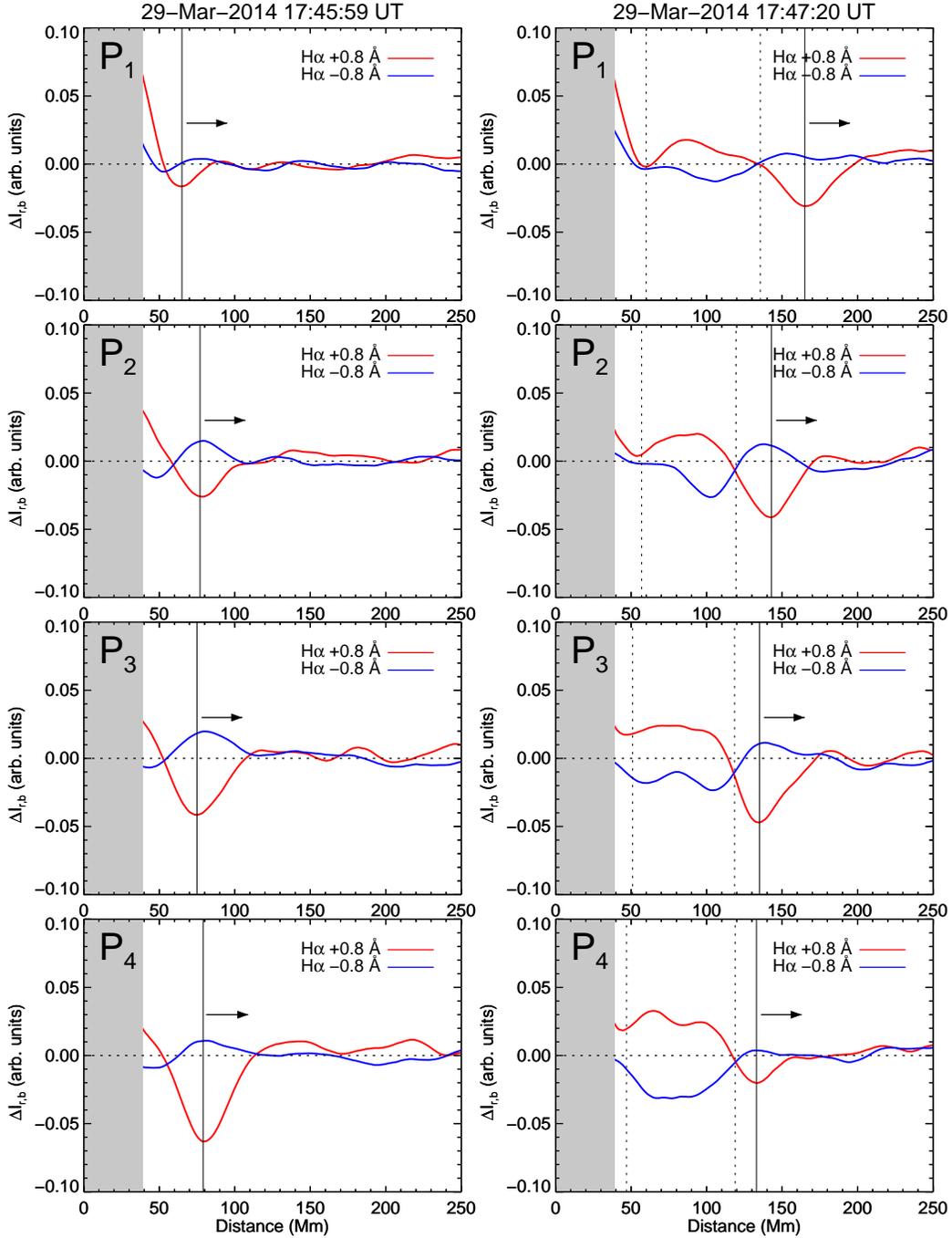}
	\caption{ {\small H$\alpha$ intensity profiles relative to the quiet-conditions at $+0.8$~{\AA} (red-wing)
	and $-0.8$~{\AA} (blue-wing), computed along trajectories P$_1$--P$_4$ at 17:45:59 and 17:47:20~UT,
	respectively. The gray area in each panel covers the contaminated signal by the scattered light of the flare.
	The profiles show perturbations as depression and enhanced patterns caused by the arrival of the coronal wave
	to the chromosphere (vertical solid lines). The profiles at $+0.8$~{\AA} exhibit a strong dip (red-ward),
	while at the same time the profiles at $-0.8$~{\AA} is slightly enhanced (blue-ward). This intensity changes
	is an indication that the H$\alpha$ spectral line is predominantly shifted red-ward.
        The intensity profiles also show subsequent disturbances to the wave passage (delimited by vertical dashed
	lines at 17:47:20~UT) showing a reversed pattern, i.e., enhancement in the red-wing and reduction of the
	intensity in the blue-wing. These delayed disturbances are related to the relaxation or recomposition state
	of the chromosphere.} }
\label{fig:FIG6}
\end{figure}

In Figure~\ref{fig:FIG6} the obtained intensity profiles at H$\alpha$ $+0.8$ and $-0.8$~{\AA}, computed along the
four paths (P$_1$--P$_4$) for the time steps 17:45:59 and 17:47:20~UT are presented. The profiles reveal perturbations
as depression and enhanced patterns caused by the arrival of the coronal wave to the chromosphere.
This manifestation which is observed simultaneously as a strong depression in absorption in the red-wing and as a
moderated enhancement in the blue-wing, corresponds to the downward motion of the chromospheric material caused by
the collision between the coronal wave moving downward with the uppermost chromospheric layer.
The characteristics observed in the profiles of H$\alpha$ wing suggests that the H$\alpha$ spectral line is
predominantly shifted red-ward in the initial response. In Figure~\ref{fig:FIG6} it is also interesting to note that
right after the downward motion the disturbed chromospheric layer exhibits a reversed pattern, i.e., brightening in
the red-wing and darkening in the blue-wing (panels at 17:47:20~UT). This is because a restoration process takes place
in the chromosphere after the plasma was forced to move downward, resulting in such a down-up swing disturbance.

\subsection{Velocity amplitude}
We further examine the strength of the disturbance in the chromosphere by computing the velocity amplitude
of the chromospheric plasma. The combination of the normalized intensity maps performed previously allow us to
produce Doppler signals. The following expression is applied to obtain the Doppler signal

\begin{equation}
	DS = \frac{I_{r}(x,t)-I_{b}(x,t)}{I_{r}(x,t)+I_{b}(x,t)},
\end{equation}

\noindent where $I_{r}$ and $I_{b}$ are H$\alpha$ intensity recorded at $+0.8$~(red-wing) and $-0.8$~{\AA} (blue-wing)
from the line center, respectively. Note that a positive Doppler signal corresponds to a blueshift (upward motion), and
it is opposite to the convention.

\begin{figure}
\centering
\includegraphics[scale=0.95]{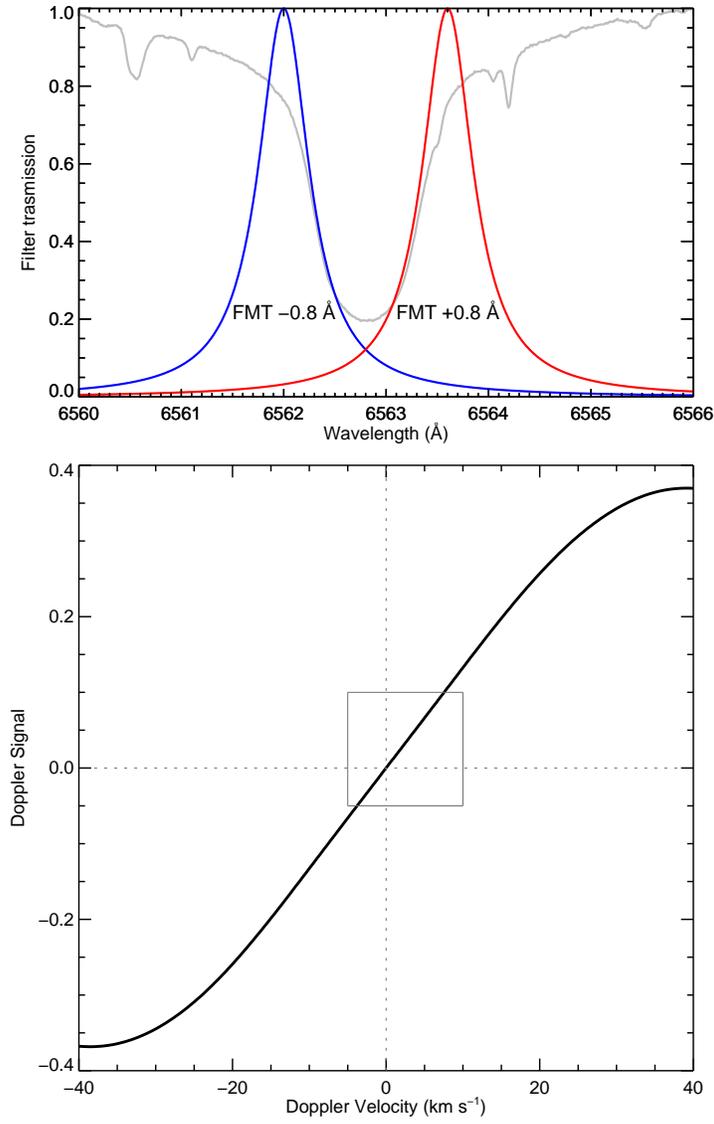}
        \caption{ {\small Upper: FMT filter-transmission profiles centered at $6562.0$~{\AA} (blue) and $6563.6$~{\AA}
	(red), respectively, both with a nominal FWHM of $0.6$~{\AA}. The gray background profile is the atlas solar
	spectrum of the H$\alpha$ line normalized to the continuum.
        Lower: synthetic Doppler signal as a result of convolution of the FMT filter-transmission profiles with the
	solar spectrum. The inset box in the plot outlines the range of the Doppler signal obtained from our observation.
	} }
\label{fig:FIG7}
\end{figure}

To quantify the Doppler signal obtained along paths P$_{1}$--P$_{4}$, we correlate them with a synthetic Doppler signal
resulting from the convolution of the H$\alpha$ solar spectrum with the FMT filter-transmission profiles centered at
$6562.0$ and $6563.6$~{\AA}.
Since the synthetic Doppler signal provides insights of the amount of wavelength shift $\Delta\lambda$ with respect
to the core wavelength $\lambda_{0}$, we can have a direct determination of the Doppler velocity 
$v = (\Delta\lambda/\lambda_{0})c$, where $c$ is the speed of light.
Representative profiles of the FMT filter-transmission together with the profile of the atlas solar spectrum are shown
in Figure~\ref{fig:FIG7}. Relation between the line-of-sight velocity and the Doppler signal is also shown in the lower
panel of Figure~\ref{fig:FIG7}.

\begin{figure}
\centering
\includegraphics[scale=0.92]{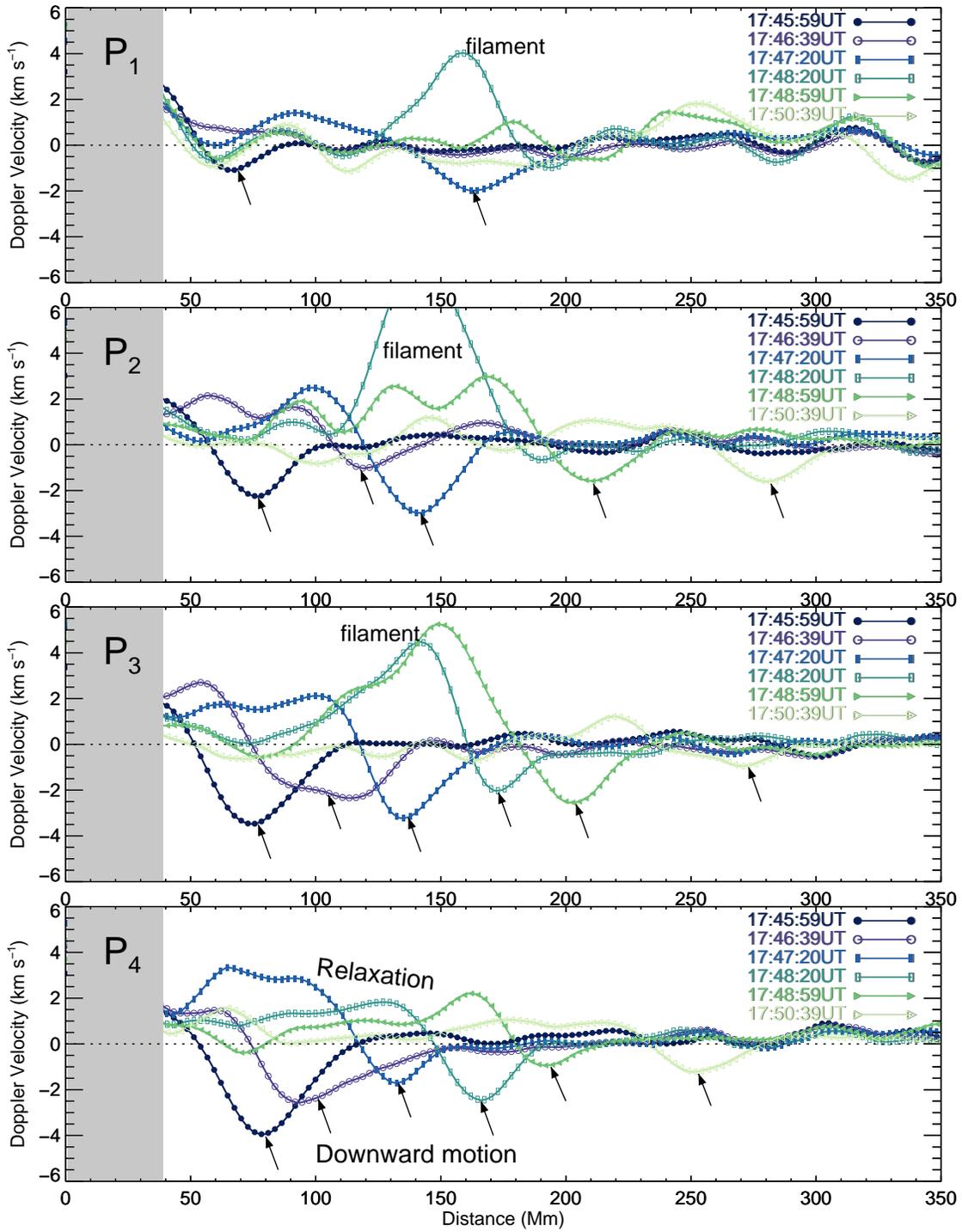}
	\caption{ {\small Doppler velocity of the Moreton wave on 2014 March 29 for six time steps along trajectories
	P$_{1}$--P$_{4}$. The velocity was estimated by correlating the obtained perturbation profiles with a synthetic
	Doppler signal (see the text). 
	The profiles show the velocity amplitude of the moving mass at the front of the Moreton wave, wherein the arrows
	point-out the downward motion of the chromospheric plasma produced by the global disturbance in the corona.
	Also observed is a subsequent relaxation process of the chromospheric layer.
	At about 17:48:20 and 17:48:59~UT the profiles also exhibit strong steepness (upward motion) along paths
	P$_{1}$, P$_{2}$, and P$_{3}$}, these features are due to the wave interaction with quiescent filaments
	standing in the propagation direction of the Moreton wave.}
\label{fig:FIG8}
\end{figure}

In Figure~\ref{fig:FIG8} we show the estimated Doppler velocity of the Moreton wave expressed as perturbation profiles,
obtained along four trajectories (P$_{1}$--P$_{4}$) for six different time steps. The perturbation profiles show a clear
evolution of the wave front progression, revealing downward motion and a subsequent relaxation process as upward motion
of the local chromospheric plasma. This represents the velocity amplitude of the moving mass at the front of the Moreton
wave. Carrying out an inspection to the perturbation profiles, we see a successive depletion (downward velocity) along
P$_{2}$, P$_{3}$, and P$_{4}$ directions for the time interval between 17:45--17:50~UT, meaning that the chromospheric
material is being pushed downward as the wave front hits and expands through the chromosphere.
The maximum depletion caused by the wave front is observed along trajectory P$_{4}$ at 14:45:59~UT, it occurs relatively
close to the source region (i.e, flare site) at a distance of $\sim$$80~{\rm Mm}$, reaching a highest value of about
$-4~\rm{km~s}^{-1}$. At the same time and distance the perturbation profiles along P$_{2}$ and P$_{3}$ show that the
plasma moves downward with a velocity of $-2.4$ and $-3.5~\rm{km~s}^{-1}$, respectively, whereas at this time the effect
of the wave front along P$_{1}$ is less pronounced. Another strongest signature of plasma compression is noticed at
17:47:20~UT in P$_{2}$ and P$_{3}$ directions located at $\sim$$140~{\rm Mm}$ from the flare site with velocities of $-2.9$
and $-3.3~\rm{km~s}^{-1}$, respectively, and at 17:48:20 and 17:48:59~UT along trajectories P$_{4}$ and P$_{3}$ having
velocities of about $-2.5$ and $-2.7~\rm{km~s}^{-1}$, respectively.
It is noted that the amplitude of the velocity profiles decreases with increasing distance. For example, this effect is
much more evident along P$_{2}$, P$_{3}$ and P$_{4}$ directions, suggesting that the strength of the large-amplitude 
coronal disturbance weakens as the time goes on.

The perturbation profiles in Figure~\ref{fig:FIG8} also exhibit strong steepness (upward motion) along paths P$_{1}$,
P$_{2}$, and P$_{3}$, particularly at 17:48:20 and at 17:48:59~UT. These features are not due to the response of the
local chromospheric plasma to the wave front passage, rather, due to the interaction with quiescent filaments laying
down in the propagation direction of the Moreton wave.

\subsection{Velocity map}
\begin{figure}
\centering
\includegraphics[scale=0.82]{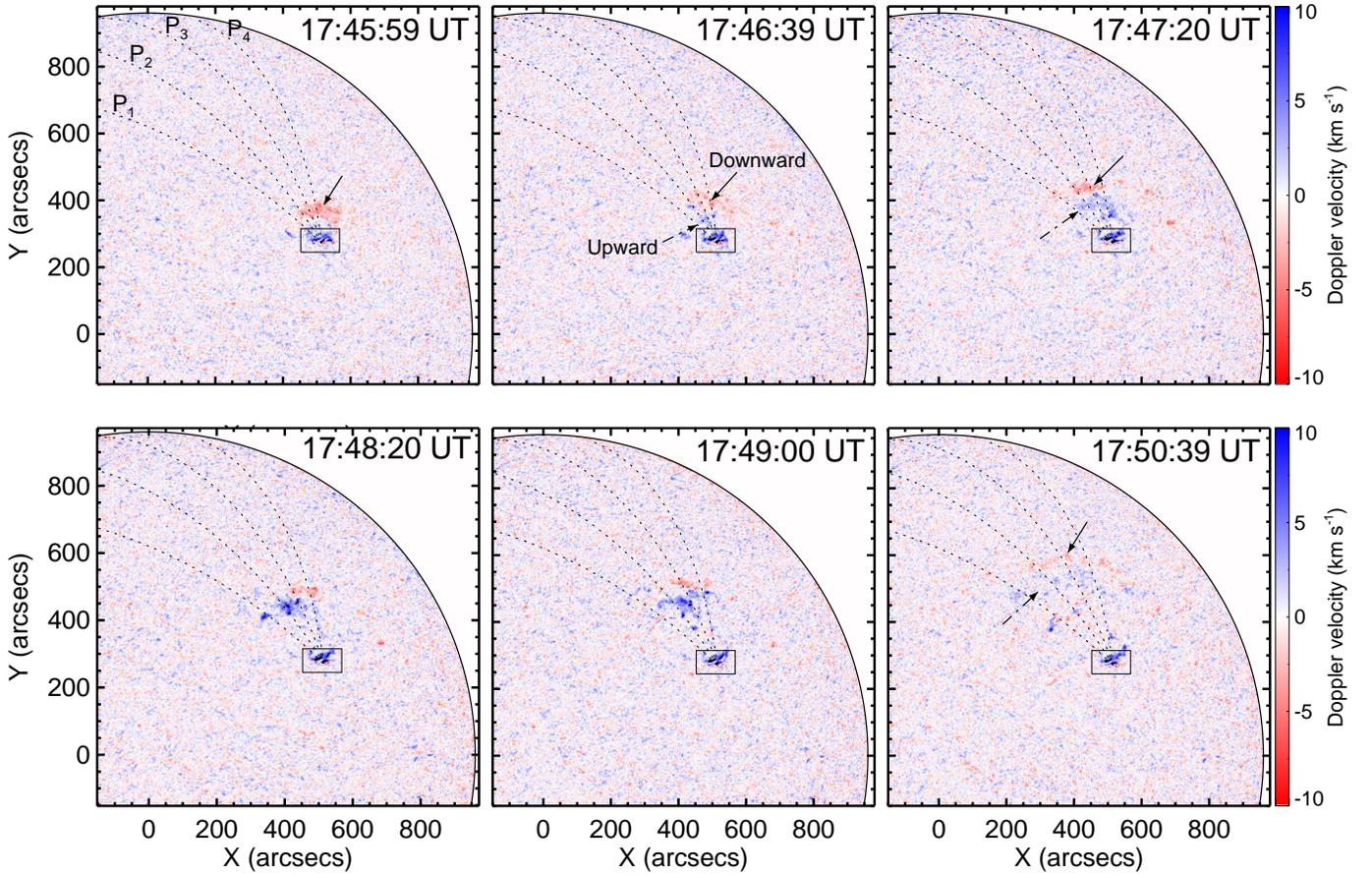}
        \caption{ {\small Doppler maps of the Moreton wave on 2014 March 29. The time period is the same to that shown in
	the perturbation profiles (Figure~\ref{fig:FIG8}). The wave fronts highlighted in red and blue correspond to downward
	and upward motions of the chromospheric plasma (see the text). The leading edge of the Moreton wave observed in red
	may be related to plasma compression, while the trailing seen in blue is an indication of recovery or relaxation
	process of the chromosphere. The rectangles in each panel depict the flare site, and the dashed lines outlines the
	trajectories P$_{1}$--P$_{4}$ considered in our analysis. 
	The online animated figure shows the Doppler characteristics of the Moreton wave from 17:44 to 18:00~UT.} } 
\label{fig:FIG9}
\end{figure}

The intensity maps described in section~4.1 are also used to characterize the mass motion of the local chromospheric
material caused by the Moreton wave progression. Applying equation~(1) we produced Doppler signal maps, then
base-difference maps are obtained by subtracting a pre-event map taken at 17:42:39~UT. The subtraction essentially
permit us to preserve the wave front signatures as residuals, as well as to enhance the wave contrast.
Following the procedure to determine the velocity amplitude in section~4.2, by cross-correlating the base-difference
maps with the synthetic Doppler signal, the line-of-sight velocity is calculated at every pixel of the map.
The results are presented in Figure~\ref{fig:FIG9} as time-series Doppler maps, wherein it is revealed that the
leading edge of the traveling wave~front appears as downward motion of the chromospheric structure, followed closely
by a much wider disturbance seen as the upward motion due to the restoration of the chromospheric layer. It can be
noted that the wave fronts develop semicircular shapes, showing a preferential propagation direction north-east from
its origin, possibly moving toward the regions of low Alfv\'en speed distribution (see discussion in section~6.1). 
As in the perturbation profiles (Figure~\ref{fig:FIG8}), high downward velocity is observed in the initial stage of
the wave propagation. It is also observed that the Moreton wave becomes progressively more diffuse and irregular,
the wave signature is no longer recognized after 17:53:20~UT (see the animation of Figure~\ref{fig:FIG9}).

\section{Coronal Shock Diagnostic}
In this section we focus on the characteristics of the coronal shock wave and its interaction with the underlying
chromosphere. We explore the scenario in which the Moreton wave is created under the action of the MHD fast-mode
wave propagation in the corona.

\subsection{Corona-chromosphere interaction}
In Figure~\ref{fig:FIG2}c it is noted that the footprint of the dome-like expanding coronal wave morphologically
and spatially corresponds to the wave front observed in He~{\sc ii} and H$\alpha$ lines (Figure~\ref{fig:FIG2}c--f).
The manifestation of the local plasma in these cooler lines is in response to the compression exerted by the
downward-propagating disturbance in the corona, such as discussed in previous sections.

\begin{figure}
\centering
\includegraphics[scale=0.55]{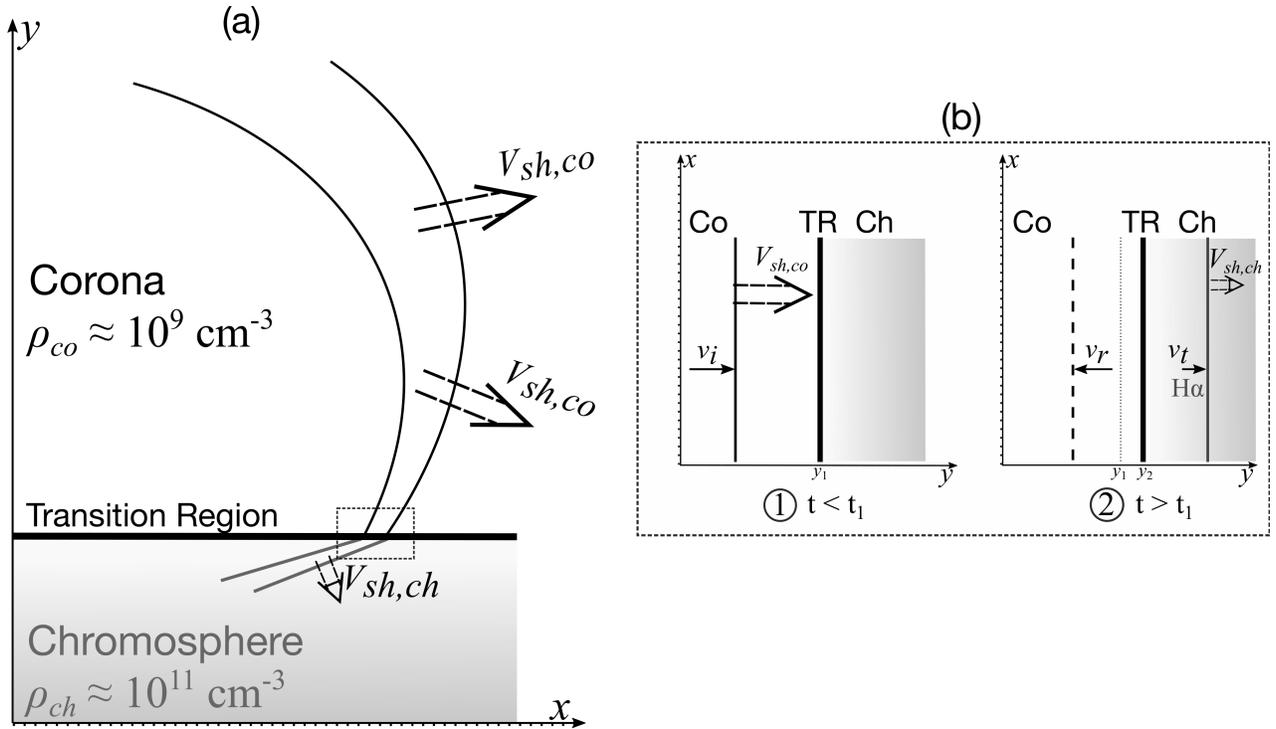}
	\caption{ {\small (a) Schematic picture of a globally propagating shock wave in the corona $V_{sh,co}$ 
	intersecting the transition region and the chromosphere $V_{sh,ch}$. (b) Enlarged view of the interaction
        corona--chromosphere (dotted rectangle in panel (a), note the change of axis orientation). The illustration
	shows two instants of interaction: (1) shock wave propagating in the corona just before the collision
	with the transition region ($t<t_{1}$), and (2) subsequent period ($t>t_{1}$).
	At the time when the shock wave reaches the transition region, the downward motion of the plasma traveling
	behind the shock with a velocity $v_{i}$ (incident wave) leads to a downward shift the transition region
	from $y_{1}$ to $y_{2}$, this happens at $t>t_{1}$. As a consequence, the chromosphere is forced to move
	downward with a velocity $v_{t}$, this is the velocity amplitude of the transmitted wave from the corona to
	the chromosphere right after the collision and can be regarded as the velocity amplitude of the Moreton wave
	seen in H$\alpha$ line.
	Additionally, due to the density discontinuity at the boundary of corona--chromosphere, a reflected wave
	$v_{r}$ is also identified traveling in opposite direction in the corona.} }
\label{fig:FIG10}
\end{figure}

The interaction of the corona and chromosphere can be treated as a discontinuity problem. In Figure~\ref{fig:FIG10}a
a schematic picture of a globally propagating shock in the corona intersecting the transition region and the
chromosphere is presented. The illustration also shows two instants of interaction (Figure~\ref{fig:FIG10}b), namely
shock wave propagating in the corona just before the collision with the transition region ($t<t_{1}$), and the
subsequent period ($t>t_{1}$).
In this scheme the transition region acts as the interface layer between the corona and chromosphere, which in our
analysis represents the contact discontinuity. Since the transition region is a thin layer, it is expected strong
effects because the collision. At the time of shock arrival on the transition region, the downward motion of the
plasma traveling behind the shock with a velocity $v_{i}$, this being the velocity amplitude of the incident wave,
leads instantaneously to a downward motion of the local plasma pushing down the transition region and setting the
motion in the chromosphere with a velocity $v_{t}$. This latter represents the velocity amplitude of the transmitted
wave from the corona to the chromosphere right after the collision, and can be interpreted as the velocity amplitude
of the chromosphere observed in H$\alpha$ line during the passage of the Moreton wave. In addition, because the sudden
density jump between the corona and chromosphere, a reflected wave traveling in the corona in opposite direction is
also expected and the velocity of the upward-moving plasma is denoted by $v_{r}$ in our analysis
(see Figure~\ref{fig:FIG10}b).

In one-dimensional discontinuity problem and with the aid of the conservation laws of momentum flux and energy
flux, the shock wave transmission from the corona into the chromosphere can be related as follows

\begin{equation}
        \rho_{co}(v_{i} - v_{r})c_{co} = \rho_{ch}v_{t}c_{ch}
\end{equation}

\begin{equation}
        \rho_{co}(v_{i}^{2} - v_{r}^{2})c_{co} = \rho_{ch}v_{t}^{2}c_{ch}.
\end{equation}

\noindent In the above equations, $v$ is the velocity amplitude of the wave, where the subscripts denote the
incident $i$, reflected $r$, and transmitted $t$ waves, as discussed above. The plasma density and the phase
velocity of the MHD fast-mode wave in the corona $co$ and chromosphere $ch$ are denoted by $\rho_{co}$,
$\rho_{ch}$ and $c_{co}$, $c_{ch}$, respectively. Similar to~\citet{takahashi2015} from equations~(2) and (3),
it is found a direct relation between the velocity amplitude of incident and transmitted waves

\begin{equation}
        v_{i} = \left(\frac{1 + \sqrt{a}}{2}\right)v_{t},
\end{equation}

\noindent here $a = \rho_{ch}/\rho_{co}$ is the density ratio of chromosphere and corona. The velocity amplitude we
derived from H$\alpha$ observation (Figure~\ref{fig:FIG8}) is a direct indicator of $v_{t}$.
The result of the velocity amplitude of incident wave $v_{i}$ calculated with equation~(4) is listed in Table~{2}.
The calculation is done assuming the density in the chromosphere and the corona to be $\rho_{ch}\approx10^{11}$ and
$\rho_{co}\approx10^{9}\rm{cm^{-3}}$, respectively, and also taking as $v_{t}$ the highest downward velocity observed
in the perturbation profiles of the Moreton wave.

\begin{table}[!ht]
\setcounter{table}{2}
\tablenum{2}
\centering
\caption{Velocity amplitude of incident $v_{i}$ and transmitted $v_{t}$ waves along trajectories
        P$_{1}$--P$_{4}$.}
\begin{tabular}{lcclcclcclcc}
        \hline
        \hline
        Time (UT)       &\multicolumn{2}{c}{P$_{1}$} &  &\multicolumn{2}{c}{P$_{2}$} &  &\multicolumn{2}{c}{P$_{3}$} &  &\multicolumn{2}{c}{P$_{4}$}\\
\cline{2-3}\cline{5-6}\cline{8-9}\cline{11-12}
                        &$v_{i}$    &$v_{t}$    &       &$v_{i}$ &$v_{t}$       &       &$v_{i}$ &$v_{t}$       &       &$v_{i}$ &$v_{t}$\\
\hline
        17:45:39        &6.05      &1.1         &       &6.05   &1.1            &       &13.7   &2.5            &       &15.9   &2.9    \\
        17:45:59        &6.60      &1.2         &       &13.2   &2.4            &       &19.2   &3.5            &       &22.0   &4.0    \\
        17:46:20        &6.05      &1.1         &       &8.25   &1.5            &       &11.0   &2.0            &       &15.9   &2.9    \\
        17:46:39        &--        &--          &       &6.05   &1.1            &       &13.2   &2.4            &       &13.7   &2.5    \\
        17:46:59        &--        &--          &       &8.80   &1.6            &       &18.1   &3.3            &       &11.0   &2.0    \\
        17:47:20        &11.0      &2.0         &       &15.9   &2.9            &       &18.1   &3.3            &       &10.4   &1.9    \\
        17:47:39        &8.80      &1.6         &       &11.0   &2.0            &       &20.35  &3.7            &       &8.25   &1.5    \\
        17:47:59        &7.69      &1.4         &       &9.90   &1.8            &       &19.8   &3.6            &       &13.2   &2.4    \\
        17:48:20        &5.50      &1.0         &       &--     &--             &       &12.6   &2.3            &       &13.7   &2.5    \\
        17:48:39        &--        &--          &       &5.50   &1.0            &       &8.25   &1.5            &       &11.0   &2.0    \\
        17:49:00        &--        &--          &       &--     &--            	&       &14.8   &2.7            &       &5.50   &1.0   \\
        17:49:20        &7.69      &1.4         &       &9.90   &1.8            &       &11.0   &2.0            &       &4.40   &0.8    \\
        17:49:39        &--        &--          &       &--   	&--            	&       &13.2   &2.4            &       &3.30   &0.6    \\
        17:50:00        &11.0      &2.0         &       &3.30   &0.6            &       &12.6   &2.3            &       &3.30   &0.6    \\
        17:50:20        &--        &--          &       &--   	&--            	&       &10.4   &1.9            &       &7.15   &1.3    \\
        17:50:39        &--        &--          &       &8.80   &1.6            &       &5.50   &1.0            &       &6.60   &1.2    \\
        17:50:59        &--        &--          &       &8.25   &1.5            &       &3.85   &0.7            &       &6.05   &1.1    \\
\hline
\end{tabular}\label{tab:first}%
	\tablecomments{($v_{i}$ and $v_{t}$ in units of ${\rm km~s^{-1}}$). The blanks indicate the time where it is not possible
	to determine signatures of downward motion in the H$\alpha$ perturbation profiles.}
\end{table}

\subsection{Shock strength in the corona based on the Moreton wave velocity amplitude}

The purpose of this section is to show how our H$\alpha$ results can further be used to characterize the strength
of the incident shock in the corona.
By using the Rankine-Hugoniot relations and taking advantage of the weak shock approximation which states that
the discontinuity in every quantity is small \cite[cf.][]{landau1987}, that is, the shock speed is slightly larger than the
sound speed, it is possible to relate the velocity amplitude of the shock in the corona to that of the velocity amplitude of
the disturbance in the chromosphere, the latter being the H$\alpha$ velocity amplitude of the Moreton wave. The following
expression summarize the above statements

\begin{equation}
        v_{t} = -\frac{4}{\gamma+1}(M^{2}_{i} -1)\frac{c_{ch}}{1 +c_{ch}/c_{co}},
\end{equation}
where $v_{t}$ is the velocity amplitude of the disturbance in the chromosphere, $M_{i}$ the Mach number of the incident shock
in the corona, and $c_{ch}$, $c_{co}$ are the propagation speed of the shock in the chromosphere and corona, respectively.
The derivation of this equation is presented in the Appendix~A (see also Figure~\ref{fig:FIG15}).
Equation~(5) allows to estimate the Mach number of the shock in the corona for a given value of the H$\alpha$ velocity
amplitude.

Let us discuss equation~(5) in more detail. Under the weak shock approximation, the propagation speed of the shock in the
corona is the sound speed, under this condition the situation in the chromosphere may be also the same. 
For example, for a typical temperature in the chromosphere $T\approx10^{4}~{\rm K}$, one finds the sound speed
$c_{s}=\sqrt{(\gamma~k~T)/m_{p}} \approx10~{\rm km~s^{-1}}$, where $k$ is the Boltzmann constant, $m_{p}$ the proton mass,
and $\gamma$ the adiabatic index of the gas. In weak shock events this sound speed would be comparable to the speed of the shock
in the chromosphere, as we noted above.
Furthermore, because the sound speed in the corona is about one order of magnitude larger than that in the chromosphere, or
even more, the term $c_{ch}/c_{co}$ in the equation can be neglected.

Taking into account the above considerations and making use of the derived velocity amplitude of the Moreton wave $v_{t}$
along with the adiabatic index $\gamma=5/3$, equation~(5) can be now evaluated. The results are shown in Table~3.
We stress that the calculation presented here is in one-dimensional regime and a pure hydrodynamic approach, no magnetic field
effect is being considered. A future work will be devoted on this calculation including much more details.

\begin{deluxetable}{lcccc}[h!]
	\tablecaption{Mach number of the incident shock in corona along trajectories
        P$_{1}$--P$_{4}$, based on the Moreton wave velocity amplitude. \label{tab:mathmode}}
\tablecolumns{5}
\tablenum{3}
\tablewidth{0pt}
\tablehead{ \colhead{Time (UT)} & \colhead{P$_{1}$} & \colhead{P$_{2}$} & \colhead{P$_{3}$} & \colhead{P$_{4}$} }
\startdata
17:45:39  &1.04  &1.04  &1.08  &1.09 \\
17:45:59  &1.04  &1.08  &1.11  &1.12 \\
17:46:20  &1.04  &1.05  &1.06  &1.09 \\
17:46:39  &--    &1.04  &1.08  &1.08 \\
17:46:59  &--    &1.05  &1.10  &1.06 \\
17:47:20  &1.06  &1.10  &1.11  &1.06 \\
17:47:39  &1.05  &1.07  &1.11  &1.05 \\
17:47:59  &1.05  &1.06  &1.11  &1.08 \\
17:48:20  &1.03  &--    &1.07  &1.08 \\
17:48:39  &--    &1.03  &1.05  &1.06 \\
17:49:00  &--    &--    &1.08  &1.03 \\
17:49:20  &1.04  &1.06  &1.06  &1.03 \\
17:49:39  &--    &--    &1.08  &1.02 \\
17:50:00  &1.06  &1.02  &1.07  &1.02 \\
17:50:20  &--    &--    &1.06  &1.04 \\
17:50:39  &--    &1.05  &1.03  &1.04 \\
17:50:59  &--    &1.05  &1.02  &1.03 \\
\hline
\enddata
\tablecomments{The blanks indicate the time where it is not possible to determine
        signatures of downward motion in the H$\alpha$ perturbation profiles.}
\end{deluxetable}

\subsection{Shock characteristics in the corona}
According to the MHD shock theory the jump relation that describes an oblique shock is defined as~\citep{priest2000}

\begin{eqnarray}\nonumber
(v_{1}^{2}-X v_{A1}^{2})^{2}\{ X c_{s1}^{2}+\frac{1}{2}v_{1}^{2}\cos^2\theta(X(\gamma-1)-(\gamma+1)) \}\\ \nonumber
        +\frac{1}{2}v_{A1}^{2}v_{1}^{2}\sin^2\theta~X\\ \nonumber
	\times\{ (\gamma+X(2-\gamma))v_{1}^{2}-X v_{A1}^{2}((\gamma+1)-X(\gamma-1) ) \}\\
	=0,~~~~
\end{eqnarray}

\noindent where $X \equiv\rho_2/\rho_1$ is the compression ratio or density jump at the front of the shock, $\theta$
is the angle between the upstream magnetic field and the shock normal, $c_{s}$ and $v_{A}$ are the sound and Alfv\'en
speeds, respectively. Taking for the adiabatic index $\gamma=5/3$ and the sound speed $c_{s}^{2}=(\gamma\beta/2)v_{A}^{2}$,
equation~(6) can be rewritten in terms of the Alfv\'en Mach number $M_{A}$ as follows \cite[see the Appendix~B, also]
[]{vrsnak2002b}

\begin{eqnarray}\nonumber
        (M_{A1}^{2}-X)^2\{ 5\beta X+2M_{A1}^{2}\cos^2\theta(X-4) \} + M_{A1}^{2}\sin^2\theta~X\\
        \times\{ M_{A1}^2(5+X)+2X(X-4) \}=0.~~~~
\end{eqnarray}

\noindent Equation~(7) depends on the inclination $\theta$, the compression ratio $X$, and also on the ratio of the
plasma to the magnetic field pressure $\beta$. Under the assumption that the proton number density $n$ of the coronal
emitting plasma is constant along the line-of-sight, the compression ratio may roughly be estimated from the emission
measure $EM\simeq n^{2}l$, where $l$ is the depth of the emitting plasma along the line-of-sight.
Since the coronal mass density is $\rho\simeq nm_{p}$, $m_{p}$ being the proton mass, it is reliable to express the
following quantities as ${EM_{2}}/{EM_{1}}={\rho_{2}^2}/{\rho_{1}^2}$, here the subscripts 1 and 2 denote the emission
measure and mass density ahead of and behind the shock front, respectively, and we assume that $l$ is common between
them, therefore we can write

\begin{equation}
        X \equiv \frac{\rho_{2}}{\rho_{1}} = \sqrt{\frac{EM_{2}}{EM_{1}}}.
\end{equation}

\begin{figure}
\centering
\includegraphics[scale=0.48]{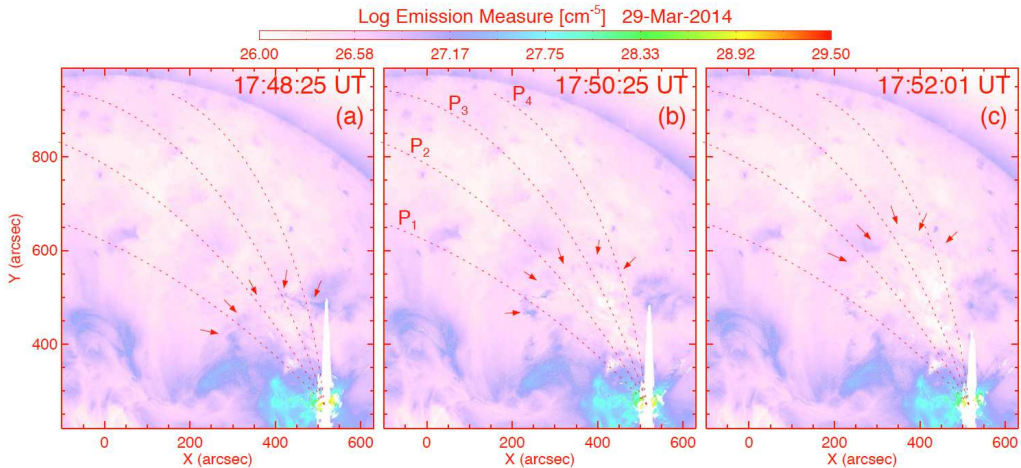}
	\caption{ {\small Differential emission measure (DEM) maps showing three instants of the coronal wave on 2014
	March 29. We used the method of \cite{cheung2015}, which enabled us to obtain DEM maps for a set of temperature
	bins. The panels show DEM solutions for a temperature range of $6.1\leq\log T/{\rm K}\leq6.4$ (see the text),
	wherein the manifestation of wave propagation (arrows) clearly can be distinguished traveling northeast.
	Based on these results, we calculate DEM perturbation profiles along trajectories P$_1$ to P$_4$ (dashed lines),
	which were used to derive the compression ratio $X$ presented in Figure~\ref{fig:FIG12}.} }
\label{fig:FIG11}
\end{figure}

\noindent The emission measure is estimated from AIA observations by applying DEM inversion method of \cite{cheung2015}.
As a first step we constructed DEM maps for a set of 21 temperature bins, spanning $\log T/{\rm K}=5.7$ to $7.7$.
We notice that DEM solutions for $\log T/{\rm K}<6.1$ do not show clearly signatures of the wave front progression, and
although solutions for $\log T/{\rm K}>6.4$ exhibit signs of the wave front, it is greatly influenced by the scattered light
of the flare. Since our interest is to trace the wave propagation, we restrict the temperature grid to
$6.1\leq\log T/{\rm K}\leq6.4$, at which the coronal wave front in DEM maps is best observed (see Figure~\ref{fig:FIG11}).
Next, considering four trajectories from the flare site (P$_{1}$--P$_{4}$), same as for the Moreton wave (see section~4),
we proceed to obtain perturbation profiles by spatially averaging the DEM solutions along each trajectory shown in
Figure~\ref{fig:FIG11}. Lastly, the computed DEM perturbation profiles in the temperature range
$6.1\leq\log T/{\rm K}\leq6.4$ were integrated to obtain the total emission measure $EM$, which enabled us to derive the
compression ratio $X$ by applying equation~(8). It should be noted that for the calculation of the compression ratio we have
considered as $EM_{1}$ in equation~(8) a pre-shocked emission measure at 17:42:37~UT. Figure~\ref{fig:FIG12} shows the
obtained compression ratio distribution along four trajectories, corresponding to the nearest time interval of the Moreton wave
velocity amplitude shown in Figure~\ref{fig:FIG8}.

On the other hand, the pre-shocked plasma temperature in the corona was constrained by applying the XRT filter ratio method
\citep{narukage2011} on two pairs of filters (Al-poly/Be-thin), which results in $T_{1}\simeq2.5{\rm~MK}$. Considering this
value the sound speed in the corona $c_{s} = \sqrt{(\gamma~k~T_{1})/m_{p}}$  is $185{\rm~km~s^{-1}}$. Moreover, based on XRT
Ti-poly observations the speed of the fast-mode wave in the corona can be assumed to be $c_{f}\simeq790~{\rm km~s^{-1}}$,
corresponding to the mean value of the speeds along trajectories P$_{1}$--P$_{4}$ (see Table~1). In the case of perpendicular
wave propagation ($\theta=90\arcdeg$), the Alfv\'en speed in the corona is defined as $v_{A} = \sqrt{c_{f}^2 - c_{s}^2}$,
which yields $v_{A} = 768~{\rm km~s^{-1}}$.
Finally, taking the obtained results of the sound speed and the Alfv\'en speed, the plasma-to-magnetic pressure ratio in the
corona is estimated as $\beta=0.06$, allowing us now to evaluate equation~(7) and find solutions for the Alfv\'{e}n Mach number
$M_{A}$ for given values of the inclination $\theta$.
Furthermore, if we assume the velocity of the shock in the corona to be $v_{sh}$, it is straightforward to relate the Alfv\'en
Mach number $M_{A}=v_{sh}/v_{A}$ along with the sound speed $c_{s}$ to the fast-mode (or fast magnetosonic) Mach number
$M_{f}=v_{sh}/c_{f}$. Again, for a perpendicular case the following relation holds

\begin{equation}
        M_{f} = \frac{M_{A}}{\sqrt{1 + \gamma\beta/2}}.
\end{equation}

In Figure~\ref{fig:FIG13} solutions of equation~(7) for the Alfv\'{e}n Mach number are presented. The calculations are done
by considering two scenarios of shock propagation: horizontal ($\theta=0\arcdeg$) and perpendicular ($\theta=90\arcdeg$)
(see the Appendix~B), and taking the highest possible values of the compression ratio profiles in the four trajectories;
those that are closely connected to the positions of the most prominent downward motion of the Moreton wave at each instance.
Figure~\ref{fig:FIG13} also compares the fast-mode Mach number estimated with equation~(9), this based on the Alfv\'{e}n
Mach number derived for the perpendicular case.

\section{Discussion}
\subsection{Morphology and propagation speed of the Moreton wave}
The Moreton wave on 2014 March 29 has brought interesting insights into our picture of chromospheric disturbances
associated with flares. Thanks to the multiwavelength coverage it was possible to reveal details of this wavelike
phenomenon happening in the chromosphere in conjunction with its coronal counterpart.
It is not surprising that the derived propagation speed of the Moreton wave is comparable to that of the propagation
speed of the coronal wave obtained from AIA and XRT observations, and the morphology of the wave front seen in the
chromosphere also shows a good correspondence in space and time with the footprints of the expanding dome-like
structure in the corona (Figure~\ref{fig:FIG2}).
This scenario further confirms that Moreton waves are the chromospheric manifestation of an MHD fast-mode shock
wave traveling through the corona, as theoretically anticipated by \citet{uchida1968}. Furthermore, the
multiwavelength time-distance diagrams presented in Figure~\ref{fig:FIG4} show that each region of the solar
atmosphere reacts and evolves quasi-simultaneously, implying that a globally propagating shock in the corona is
the triggering source of the observed disturbances.

We found that the mean velocity of the Moreton wave varies depending on the wave propagation direction
(Figure~\ref{fig:FIG5}, top panels). Similar behavior is also observed in the coronal wave propagation speed
(Figure~\ref{fig:FIG5}, bottom panels), although in the latter case the situation is much more complex due to the
3D expansion and the projection effect.
The difference of the propagation velocity along the analyzed trajectories can be attributed to the plasma condition
and the inhomogeneous topology through which the wave propagates. In the solar atmosphere both the density and the
magnetic field change considerably along the vertical direction, i.e., the density starts to decrease rapidly in the
low corona, while the magnetic field drops much faster in the outer corona. This aspect determine the ambient
Alfv\'{e}n speed, which is an important parameter to characterize wave propagation in the corona and their subsequent
effects.
The studied Moreton wave also exhibits a confined propagation direction to a certain sector, i.e., north-east from
the flare site. The reason of this confinement is because the MHD fast-mode wave cannot re-enter the chromosphere in
directions in which strong magnetic field does exist. Strong magnetic field also implies that high Alfv\'{e}n velocity
distribution dominates these regions; and because the wave front tends to propagate avoiding strong magnetic regions,
the wave energy flux concentrates in directions of low Alfv\'{e}n velocity. Therefore, the occurrence of this condition
in the low corona favors to the appearance of the Moreton wave in a restricted direction \cite[cf.][]{uchida1973}.
Indeed, the Moreton wave on 2014 March 29 was directed towards non-magnetic or weak-field regions, i.e., high latitudes
from the flare site, but not in opposite direction or towards the equatorial zone, places where high concentration of
magnetic field could exist because of the presence of magnetic structures and some active regions. \cite{zhang2011}
based on the analysis of a set of Moreton waves, also showed that the waves tend to propagate within a boundary of weak
magnetic field regions.

\subsection{Doppler velocity of the Moreton wave}
The quantitative results of the perturbation in the chromosphere presented as Doppler velocity in Figure~\ref{fig:FIG8},
provides a direct sign of the coronal disturbance exerted into the chromosphere. 
The perturbation profiles show that the arrival of the coronal wave to the chromosphere produces disturbances along the
four trajectories (P$_{1}$--P$_{4}$). However, its effect differs from one trajectory to another, and there is no a clear
correlation among them in terms of amplitude as a function of distance and time. The wave propagation along path P$_{1}$
causes a very weak disturbance, whereas in P$_{2}$, P$_{3}$, and P$_{4}$ directions the disturbances becomes gradually
much more significant. It is also noted that in the earlier stage, i.e., at 17:45:59 -- 17:47:20~UT, highest downward
velocity is observed particularly along P$_{2}$, P$_{3}$ and P$_{4}$ trajectories, later on the velocity amplitude
decreases.
The difference of the velocity amplitude we found along each trajectory may be a consequence of the non-homogeneous layer
(density and field strength) over which the wave propagates in the corona. In addition, because the 
downward component of the coronal wave interacts with a much more dense layer, i.e., the chromosphere, most of its
energy is reflected back into the corona and only a small fraction of it is able to reach deeper layers triggering disturbances
with strong effect in some localized regions, such as observed in the H$\alpha$ perturbation profiles.
Furthermore, the wave fronts seen in the Doppler maps (see Figure~\ref{fig:FIG9} and the associated movie) progressively
becomes wider and diffuse. This is indicative that the strength of the disturbance in the corona gets weaker at far
distances from the source, meaning that some dissipative process undergoes and as a result damping of the wave takes place,
similar to the perturbation profiles.
\cite{vrsnak2016} in their simulation results showed that after a certain time the velocity amplitude of the chromospheric
disturbance develops a small amplitude. The authors also reproduced the restoration process of the perturbed plasma,
consistent with our observational results.

\subsection{Characteristics of the coronal wave based on H$\alpha$ observation}
We have shown in section 5.1 that the MHD linear theory and the H$\alpha$ velocity amplitude of the Moreton wave, this
latter interpreted as the velocity amplitude $v_{t}$ of the transmitted wave from the corona to the chromosphere,
allowed us to infer the velocity amplitude $v_{i}$ of the incident wave in the corona under the assumption that the
density ratio of chromosphere and corona is $a\approx100$. It is important to emphasize that our calculation of $v_{i}$
should be understood as the associated counterpart of $v_{t}$ acting at the base of the corona. Taking the ratio of the
velocity amplitude of transmitted and incident waves, we define the velocity transmittance $T_{v}=v_{t}/v_{i}$. This
provides clues on how much fraction of the plasma velocity in the corona is transferred to the chromosphere when the shock
front surpasses a specific location of the dense chromospheric layer. 
To illustrate this point, we assume that $a$ extends from $1$ to $100$, and also we take $v_{t}=4~{\rm km~s^{-1}}$ and
$v_{i}=22~{\rm km~s^{-1}}$, corresponding to the derived values along path P$_{4}$ at 17:45:59~UT (see Table~2). It is
shown in Figure~\ref{fig:FIG14} that as $a$ increases $T_{v}$ is significantly reduced, so the effect of the density
increase, e.g., corona--chromosphere density jump, plays an important role. \cite{takahashi2015} (in their Figure~11)
also found a similar tendency of velocity transmittance from one-dimensional simulation of shock transmission to a
prominence.

From our findings we can further speculate that a wave traveling downward at the base of the corona with a velocity
$\gtrsim10~{\rm km~s^{-1}}$, this being the mean velocity amplitude of the incident wave calculated along trajectories
P$_{3}$ and P$_{4}$ in Table~2, could be able to produce a noticeable disturbance in the chromosphere.
We are aware that this is simply based on linear analytic solution, however, if the amplitude of the disturbance is so small,
such as in the limit corona--chromosphere, the linear theory is applicable.
\cite{harra2011} and \cite{veronig2011}, based on Hinode/EIS observation of a coronal wave on 2011 February 16, reported
a downward motion of the plasma in the corona with a velocity of about $20~{\rm km~s^{-1}}$. The authors argued that the
coronal wave pulse moving with such a speed was not strong enough to produce disturbances in the chromosphere. Indeed
there was no Moreton wave, but activation of filaments in H$\alpha$ line were identified associated with the referred
coronal wave \cite[cf.][]{cabezas2017}.
It should be mentioned that the reported downward motion was observed in Fe~{\sc xii} and Fe~{\sc xvi} spectral lines,
which provide plasma diagnostics of coronal emission at $\log T/K\gtrsim6.1$, whose formation region is believed to be
located at low or mainly at intermediate corona, and it is likely that plasma moving downward with $20~{\rm km~s^{-1}}$
at such a place could not effectively travel deeper to overcome the inert chromosphere and produce a Moreton wave.

\subsection{Shock strength in the corona based on DEM analysis}
\begin{figure}
\centering
\includegraphics[scale=0.92]{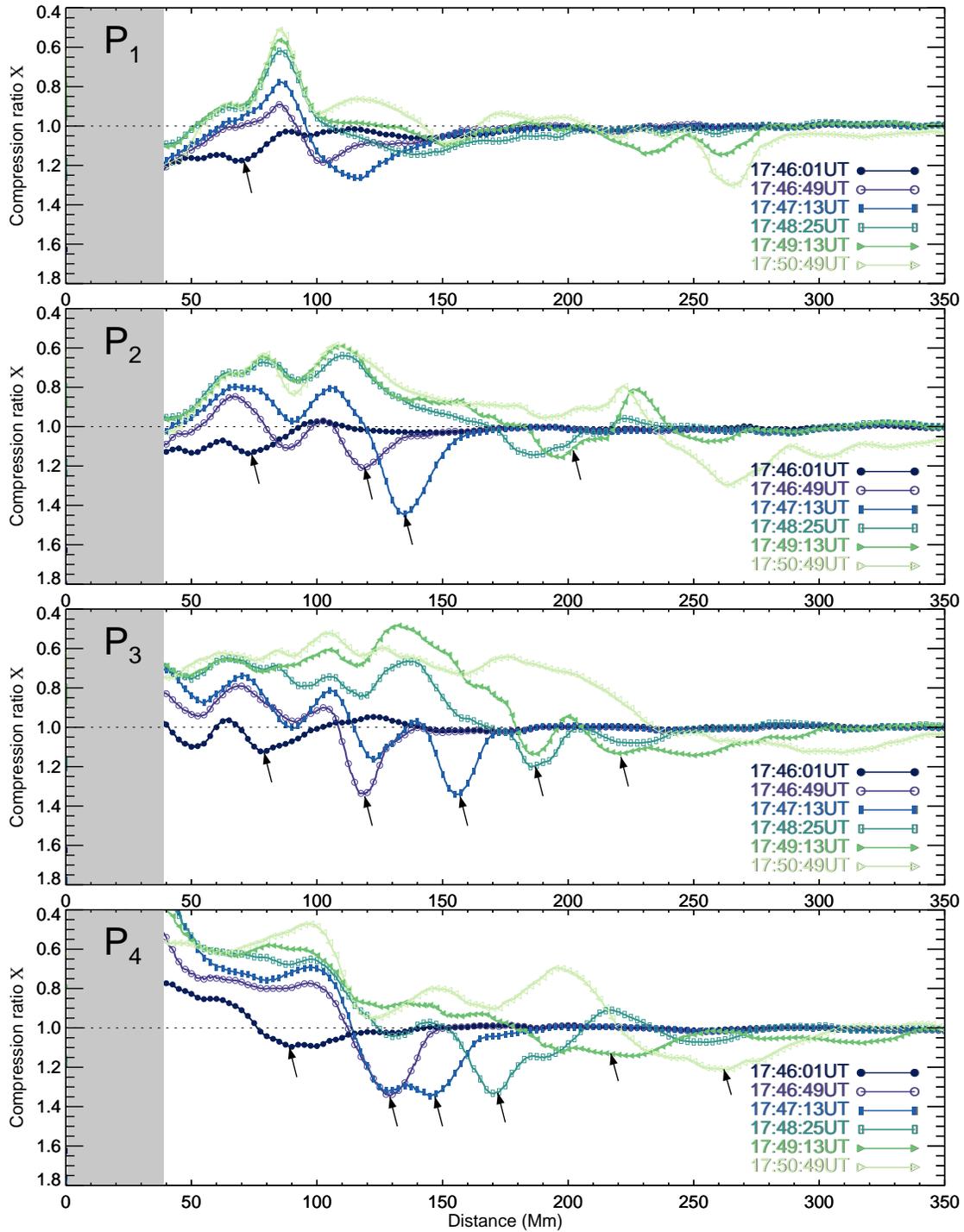}
	\caption{ {\small Compression ratio $X$ estimated from DEM maps along trajectories P$_{1}$--P$_{4}$ shown in
	Figure~\ref{fig:FIG11}. We obtained the total emission measure by integrating over a temperature range the DEM
	perturbation profiles, then equation~(8) is applied to estimate the compression ratio (see the text).
	The pre-shocked emission measure $EM_{1}$ in equation~(8) is that resulting from DEM map at 17:42:37~UT.
	The compression ratio profiles along each trajectory correspond to the nearest time interval
	of the Moreton wave velocity amplitude presented in Figure~\ref{fig:FIG8}. Note that the vertical axes are inverted,
	this for a better comparison of the observed characteristics (arrows) with those in the Moreton wave
	perturbation profiles.
	The online animation presents a synchronized time evolution of the compression ratio and the Moreton wave velocity
	amplitude along trajectory P$_{3}$, extending from 17:42 to 17:51~UT.} } 
\label{fig:FIG12}
\end{figure}

\begin{figure}
\centering
\includegraphics[scale=0.85]{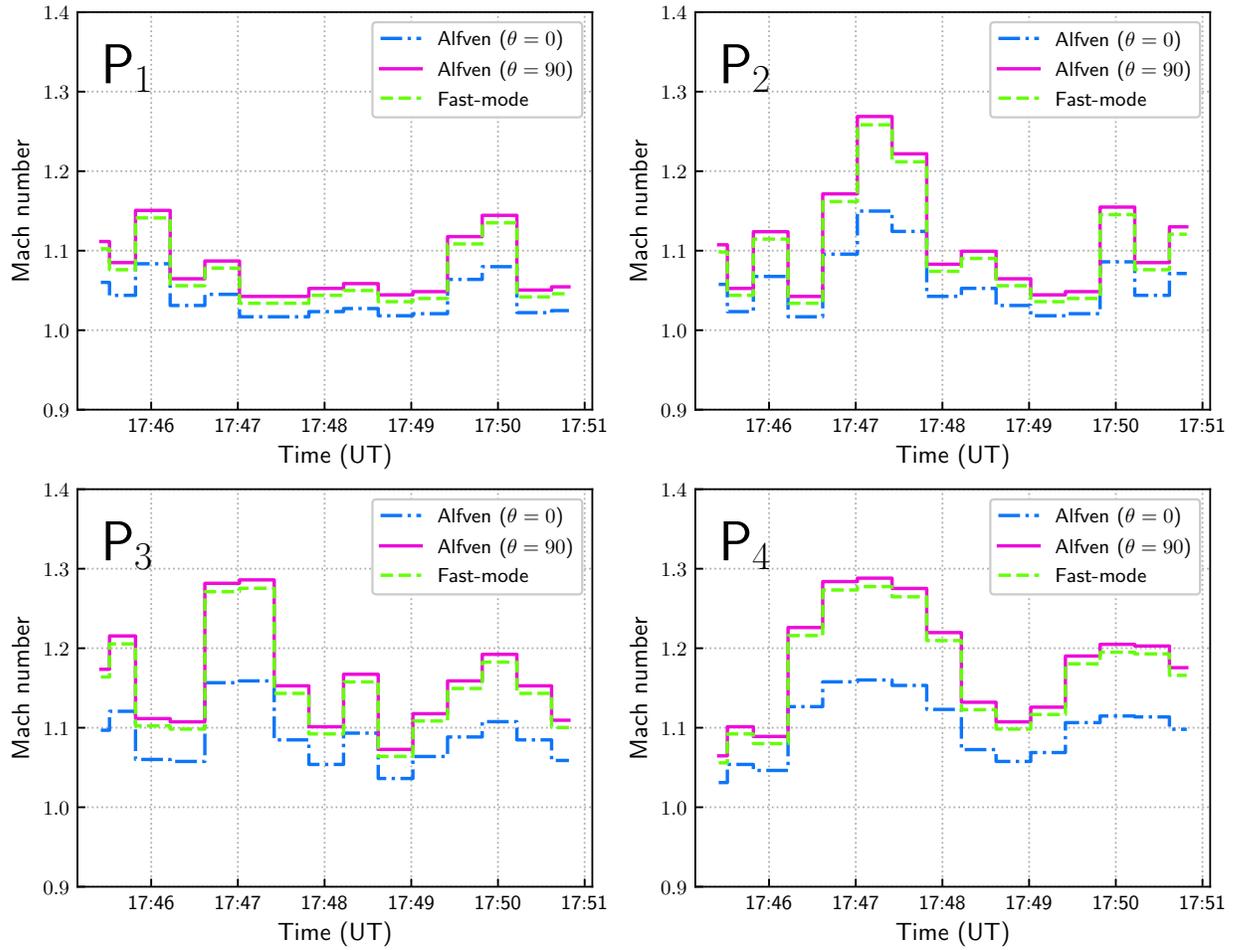}
	\caption{ {\small Alfv\'{e}n and fast-mode Mach numbers of the coronal wave on 2014 March 29 along trajectories
	P$_{1}$--P$_{4}$. The Alfv\'{e}n Mach number is based on the solution of equation~(7) using the compression
	ratio derived from differential emission measure. The profiles in blue and magenta show solutions for horizontal
	($\theta=0\arcdeg$) and perpendicular ($\theta=90\arcdeg$) shock propagation cases, respectively.
	For the calculation we take the highest values of the compression ratio profiles in each trajectory, those that
	are closely connected to the positions of large downward motion of the Moreton wave.
        On the other hand, the fast-mode Mach number (green profiles) are solutions of equation~(9) taking the results
	of the Alfv\'{e}n Mach number for the perpendicular case.} }
\label{fig:FIG13}
\end{figure}

\begin{figure}
\centering
\includegraphics[scale=0.75]{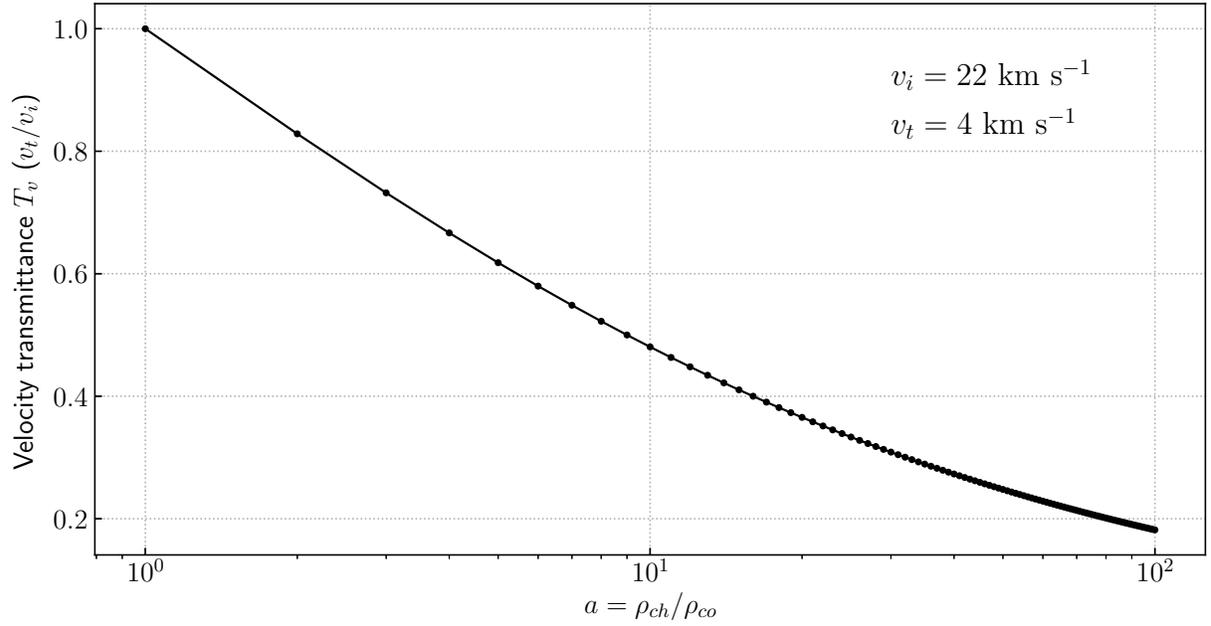}
	\caption{ {\small 
	Velocity transmittance $T_{v}$ defined as the ratio of the velocity amplitude of transmitted $v_{t}$ and incident
	$v_{i}$ waves. This provides clues on how much fraction of the plasma velocity in the corona is transferred to the
	chromosphere during the shock wave propagation. For illustration, the density jump $a$  extends from $1$ to $100$,
	and we take $v_{t}=4~{\rm km~s^{-1}}$ and $v_{i}=22~{\rm km~s^{-1}}$ corresponding to the trajectory P$_{4}$ at
	17:45:59~UT (see Table~2). In the plot it can be seen that as $a$ increases $T_{v}$ is significantly reduced.} }
\label{fig:FIG14}
\end{figure}

The compression ratio profiles shown in Figure~\ref{fig:FIG12} reveal that the coronal wave propagates by exhibiting
an irregular pattern along the different trajectories. It is found that the compression ratio, which is directly related to
the shock strength, varies non-uniformly as a function of distance and time.
On the basis of the shock wave theory, the propagation and strength of the shock in the corona is governed by the
action of the magnetic field and the ambient Alfv\'{e}n speed distribution, and in the absence of strong magnetic
field high compression ratio is expected. Therefore, the large values of the compression ratio observed along P$_2$,
P$_3$, and P$_4$ directions, implying high Mach number (see Figure~\ref{fig:FIG13}), suggests that these locations
are more likely weak magnetic field regions. Consequently, the shock traveling along these directions could produce
strong compression, resulting in relatively large amplitude chromospheric disturbance. In fact, strong downward motions
in the chromosphere are observed along the referred trajectories.
This can be seen by comparing Figures~\ref{fig:FIG8} and \ref{fig:FIG12}, in which highest values of the compression
ratio along trajectories P$_2$, P$_3$, and P$_4$, approximately match in distance and time to the large downward motion
caused by the Moreton wave (see the animation of Figure~\ref{fig:FIG12}).
Of course, a comparison between the compression ratio and the velocity amplitude of the Moreton wave cannot be done
directly, since the compression ratio based on DEM ($6.1\leq\log T/{\rm K}\leq6.4$) is a plasma diagnostic of the wave
traveling above the low corona, we are missing information of the plasma in the lowest-part of the corona. Nevertheless,
the compression ratio profiles and the velocity amplitude of the Moreton wave show a good correspondence.

The disturbance created by the wave propagation would also result in the increase of the local plasma temperature.
As we pointed out in section~5.3, DEM solutions for $\log T/{\rm K}<6.1$ do not show clear signs of the wave propagation.
In contrast, at much higher temperature range the disturbance seen as emission enhancement becomes more noticeable.
Similar findings from DEM analysis of coronal waves were also reported by \cite{vanninathan2015} and \cite{liu2018}.
The emission increasing at high temperatures is an indication of plasma heating, enhanced locally due to plasma
compression as a result of the wave propagation. The heating effect also causes intensity variation of much cooler
emission lines. For example, at AIA 171~{\AA} ($\log T$$\sim$$5.8$) coronal dimming (not shown here) is observed at the
place of the wave front as a consequence of the plasma being heated to much higher temperatures
\cite[e.g.,][]{wills-thompson1999,delannee2000,liu2010,nitta2013,liu2018}.

\section{Summary and Conclusion}
In this work we have studied in detail the Moreton wave on 2014 March 29 and its associated coronal counterpart.
We combined multiwavelength observations, including H$\alpha$ $\pm0.8$~{\AA}, He~{\sc ii} ($304$~{\AA}), coronal
emission lines from AIA, and Hinode X-ray. This data set permitted us to further understand the nature of
a globally wave propagation and their subsequent effects at different layers of the solar atmosphere. Such a global
effect is clearly evidenced in Figure~\ref{fig:FIG4}, wherein it is observed how the corona, transition region, and
the chromosphere reacted to the large-scale disturbance. It is also noted a co-spatial evolution of
the wave in the corona and the Moreton wave in the transition region and the chromosphere, particularly in the initial
stage of the propagation, thus confirming a common origin.

Thanks to the observations in the wing of the H$\alpha$ line performed by the FMT, we discussed the Doppler 
characteristics of the Moreton wave. We have shown that it is possible to quantify the downward motion of the
chromospheric material traveling at the front of the Moreton wave. The plasma at this location attains a maximum
velocity of about $4~{\rm km~s^{-1}}$, which corresponds to the velocity amplitude of chromospheric motion associated
with the Moreton wave. Our findings also showed that the chromospheric disturbance develops non-uniformly in
terms of amplitude along the analyzed trajectories (see Figure~\ref{fig:FIG8}).
Furthermore, with the aid of the MHD linear theory we also related our results of the Moreton wave velocity amplitude
with that of the plasma velocity in the corona, allowing us to infer the velocity amplitude of the incident wave from
the corona to the chromosphere. Additionally, we showed that by using the weak shock approximation in conjunction with
the velocity amplitude of the Moreton wave, the Mach number of the incident shock in corona could be estimated
and obtained a consistent result with the AIA analysis.
Very recently \cite{long2019} attempted to characterize the perturbation caused by the Moreton wave studied in
this paper, but the authors used H$\alpha$ line core observations only, in which the signature of the perturbation
cannot easily be distinguished.

The compression ratio analysis led us to conclude that the wave in the corona develops in a complex manner. This may
depend on the conditions in which the wave propagates through. As we noticed in Figure~\ref{fig:FIG12}, the compression
ratio profiles vary as a function of distance, time, and also from one trajectory to another. This aspect also has a direct
consequence in the change of the Alfv\'{e}n and fast-mode Mach numbers, such as presented in Figure~\ref{fig:FIG13}.
As for the Mach number, the values we calculated (i.e., $M_{A}\approx$1.06--1.28 and $M_{f}\approx$1.05--1.27) should be
understood as representing the strength of the shock in a confined region of the solar corona, since we restricted the compression
ratio analysis to a DEM temperature range $6.1\leq\log T/{\rm K}\leq6.4$, the missed lowest-part of the corona should be taken
into account for further conclusions.

Finally, an extension of the present work will be important especially focusing on the Doppler characteristics of the
Moreton wave, a topic which is not fully understood and scarcely explored. More studies will help us to advance and clarify
the direct connection between the downward propagating component of the coronal wave and the chromospheric reaction
leading to a Moreton wave. To this end, a combination of numerical experiments and observational results will be crucial.
As for the observational matter, it will be further constrained with the Solar Dynamics Doppler Imager
\cite[SDDI;][]{ichimoto2017}, now in operation on SMART telescope at Hida Observatory, Kyoto University, which provides
high resolution filtergrams with the spectral profiles in a wide spectral window around the H$\alpha$ line center.

\acknowledgments
{\small
The authors wish to thank the referee for constructive comments and suggestions.
We are grateful to the staff of Centro de Investigaci\'{o}n del Estudio de la Actividad Solar y sus Efectos
Sobre la Tierra, Facultad de Ciencias, Universidad Nacional San Luis Gonzaga de Ica, Per\'{u}, for providing
us the FMT data.
We thank to Drs. Petr Heinzel, Satoshi Morita, Takuya Takahashi for insightful discussions, and also to Dr.
Nariaki Nitta for valuable comments on the manuscript. {\it Hinode} is a Japanese mission developed and launched
by ISAS/JAXA, in partnership with NAOJ, NASA and STFC (UK). Additional operational support is provided by ESA
and NSC (Norway). AIA data products are courtesy of NASA/SDO science team, a mission for NASA's Living With a
Star program. DPC is supported by the Japanese government ``Monbukagakusho" scholarship from the Ministry of
Education, Culture, Sports, and Science and Technology (MEXT) of Japan. 
This work was partly supported by MEXT/JSPS KAKENHI Grant number JP15H05814, Project for Solar-Terrestrial
Environment Prediction (PSTEP). ``This work is dedicated to the memory of Dr. Mutsumi Ishitsuka (1930-2018)."}

\appendix
\section{Mach number in terms of the Moreton wave velocity amplitude}
In this section we present the derivation of equation (5) introduced in section~5.2. Here we discuss the
one-dimensional Riemann problem in hydrodynamical regime \citep{landau1987} based on the weak shock approximation.
The three-dimensionality and the magnetohydrodynamic effects are not considered.

\begin{figure}
\centering
\includegraphics[scale=0.55]{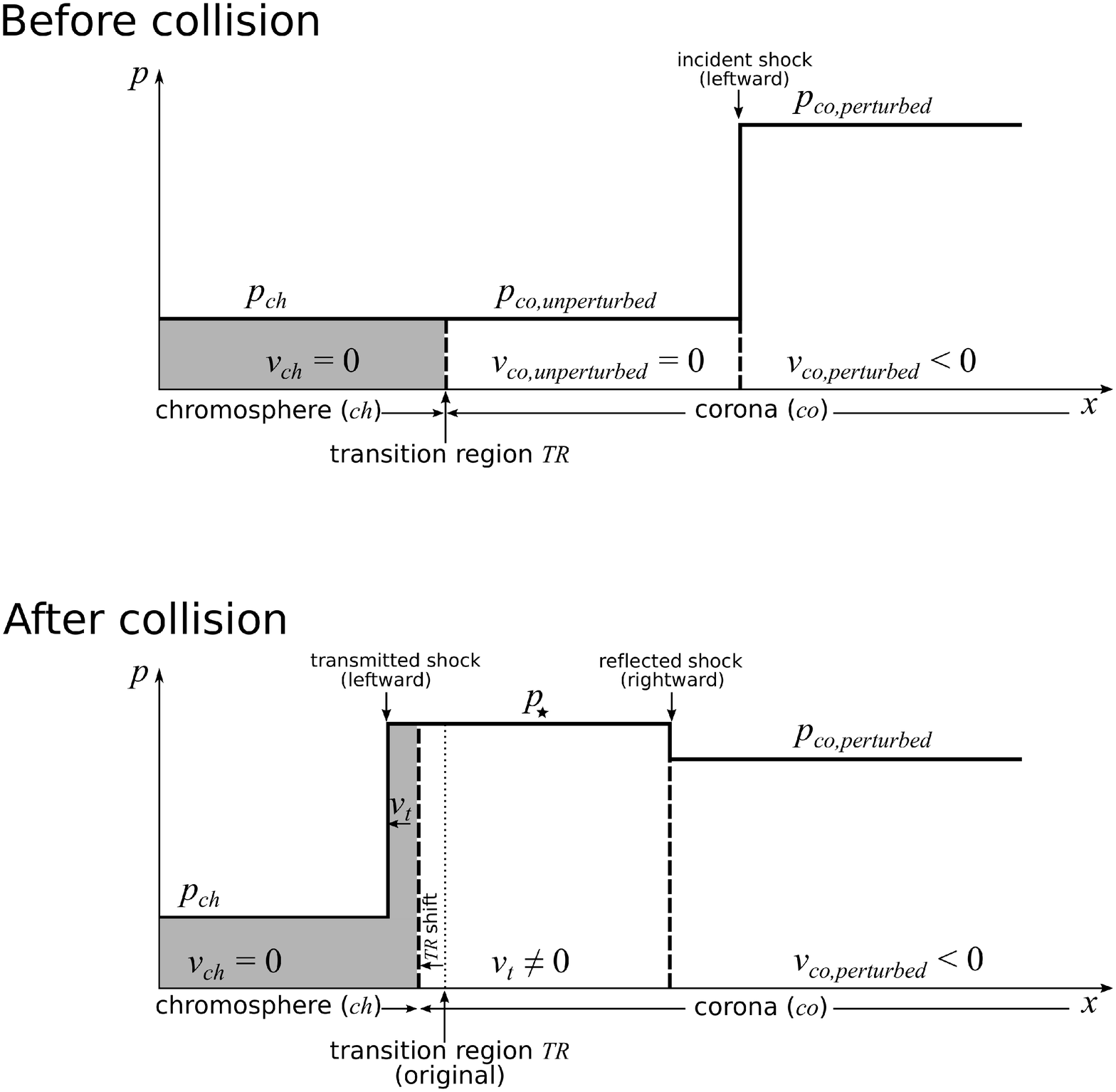}
        \caption{ {\small One-dimensional Riemann problem in hydrodynamical regime of shock interaction with the corona and
	chromosphere. In the illustration the transition region represents the contact discontinuity. The thick solid
	lines show the variation of the pressure distribution $p$ along the $x$-axis, that is, the change of the pressure in
	the corona and chromosphere before and after the collision of the shock with the contact discontinuity.} }
\label{fig:FIG15}
\end{figure}

Figure~\ref{fig:FIG15} shows an illustration of shock interaction before and after the collision with the contact
discontinuity (transition region). In the scheme the solid thick line represent the variation of the pressure $p$
along the $x$-axis. Before the collision, the initial pressure distribution in the chromosphere and corona are in
equilibrium ($p_{0}=p_{ch}=p_{co}$, $v_{ch}=v_{co} = 0$) and is separated by the contact discontinuity.
In our analysis the coronal shock (incident shock) propagates from right to left, where its downstream region is
characterized with enhanced pressure $p'_{co}=p_{co,perturbed}$ and velocity $v'_{co}=v_{co,perturbed}$.
Note that in the following we take the direction from the chromosphere to corona as positive direction, and thus,
$v'_{co}=v_{co,perturbed}<0$.
After the collision of the incident shock with the contact discontinuity, the pressure and the velocity distribution
can be divided into three regions: $p=p_{0}$, $v=0$ in the leftmost region; $p=p'_{co}$, $v=v'_{co}$ in the rightmost;
and an intermediate region characterized with $p_{\star}$ and $v_{t}$ (lower panel in Figure~\ref{fig:FIG15}).

The Rankine-Hugoniot relations tells us about the jump conditions at the incident $i$, transmitted $t$, and reflected
$r$ shocks, in which the Mach number of each shock can be expressed respectively as

\begin{equation}
        M_{i}^{2} = 1 + \frac{\gamma +1}{2\gamma}\left(\frac{p'_{co}}{p_{0}} -1\right),\quad
        M_{t}^{2} = 1 + \frac{\gamma +1}{2\gamma}\left(\frac{p_{\star}}{p_{0}} -1\right),\quad
        M_{r}^{2} = 1 + \frac{\gamma +1}{2\gamma}\left(\frac{p_{\star}}{p'_{c0}} -1\right)
\end{equation}

\begin{equation}
       0 - v'_{co} = \frac{2c_{co}}{\gamma +1}\frac{M_{i}^{2} -1}{M_{i}}, \quad
       0 - v_{t}   = \frac{2c_{ch}}{\gamma +1}\frac{M_{t}^{2} -1}{M_{t}}, \quad
      (-v'_{co}) - (-v_{t}) = \frac{2c'_{co}}{\gamma +1}\frac{M_{r}^{2} -1}{M_{r}}
\end{equation}
where $c_{co} = \sqrt{\gamma p_{0}/\rho_{co,unperturbed}}$,
	$\quad c_{ch} = \sqrt{\gamma p_{0}/\rho_{ch,unperturbed}}$,
	$\quad c'_{co} = \sqrt{\gamma p'_{0}/\rho_{co,perturbed}}$~;
therefore

\begin{equation}
	-v'_{co} = \frac{2c_{co}}{\gamma+1}\frac{M^{2}_{i}-1}{M_{i}}=\frac{2c'_{co}}{\gamma+1}
	\frac{\frac{\gamma+1}{2\gamma}\left(\frac{p_{\star}}{p'_{co}}-1\right)}
	{\sqrt{1 + \frac{\gamma+1}{2\gamma}\left(\frac{p_{\star}}{p'_{co}}-1\right)}}
	+ \frac{2c_{ch}}{\gamma+1}\frac{\frac{\gamma+1}{2\gamma}\left(\frac{p_{\star}}{p_{co}}-1\right)}
        {\sqrt{1 + \frac{\gamma+1}{2\gamma}\left(\frac{p_{\star}}{p_{co}}-1\right)}}.
\end{equation}
Because $p'_{co}$ can be described with $p_{0}$ for given $M_{i}$ in equation~(A1), the second equality of equation (A3)
determines the dependence of $p_{\star}$ on $p_{0}$ and $M_{i}$, and thus, $M_{t}$ and $v_{t}$ can be also determined as
a function of $p_{0}$ and $M_{i}$ by equation (A2).
Particularly, for weak shock
$M^2_{i}-1\ll 1~(\Leftrightarrow \delta:=p'_{co}/p_{0}-1 \ll 1)$, $\delta_{\star}:=p_{\star}/p_{0}-1 \ll 1$, then we have

\begin{equation}
	M^2_{i}-1\approx\frac{\gamma+1}{2\gamma}\delta,\quad
	\frac{\frac{p_{\star}}{p'_{co}}-1}{ \sqrt{1+\frac{\gamma+1}{2\gamma}\left(\frac{p_{\star}}{p'_{co}}-1\right) } }
	\approx \delta_{\star}-\delta,\quad
	\frac{ \frac{p_{\star}}{p_{0}}-1}{\sqrt{1+\frac{\gamma+1}{2\gamma}\left(\frac{p_{\star}}{p_{0}}-1\right)} }
	\approx\delta_{\star}.
\end{equation}
The previous statements lead us to express equation (A3) as
\begin{equation}
	\frac{c_{co}}{\gamma}\delta \approx \frac{c_{co}}{\gamma}(\delta_{\star}-\delta)+\frac{c_{ch}}{\gamma}\delta_{\star},
\end{equation}
which is reduced to
\begin{equation}
	\delta_{\star} = \frac{2}{1+c_{ch}/c_{co}}\delta.
\end{equation}
Furthermore, we can relate the Mach numbers of incident and transmitted shock as
\begin{equation}
	M^2_{t}-1 \approx \frac{2}{1+c_{ch}/c_{co}}(M^2_{i}-1),
\end{equation}
then from equations in (A2) we arrive to an expression that relates $v_{t}$ to the Mach number of transmitted and incident
shocks from the corona
\begin{equation}
	v_{t} \approx -\frac{2c_{ch}}{\gamma+1}(M^2_{t}-1)
	\approx-\frac{4c_{ch}}{\gamma+1}(M^2_{i}-1)
	\frac{1}{1+c_{ch}/c_{co}}.
\end{equation}
It should be noted again that $v_{t}$  is the velocity in the intermediate region which appears after the collision between
the shock and the contact discontinuity~(Figure~\ref{fig:FIG15}). In other words, the contact discontinuity starts to move
leftward with the velocity of $v_{t}$, affecting also the chromosphere and could be observed as the Doppler shift of H$\alpha$
line.

\section{Alfv\'{e}n Mach number from oblique shock jump relations}
Here we show the solution of equation~(6) in terms of Alfv\'{e}n Mach number. As in section~5.3 the
following notation describes an oblique shock \citep{priest2000}
\begin{eqnarray}\nonumber
(v_{1}^{2}-X v_{A1}^{2})^{2}\{ X c_{s}^{2}+\frac{1}{2}v_{1}^{2}\cos^2\theta(X(\gamma-1)-(\gamma+1)) \}
        +\frac{1}{2}v_{A1}^{2}v_{1}^{2}\sin^2\theta X\\
        \times\{ (\gamma+X(2-\gamma))v_{1}^{2}-X v_{A1}^{2}((\gamma+1)-X(\gamma-1) ) \}=0.
\end{eqnarray}
Assuming the adiabatic index $\gamma=5/3$ and expressing the sound speed as
$c^{2}_{s}=(\gamma\beta/2)v^{2}_{A}$, the first and second part of equation~(B9) can be written
respectively as

\begin{eqnarray}\nonumber
        \frac{1}{3}(v_{1}^{2} - X v^{2}_{A1})^2\left\{ \frac{5}{2}X\beta v_{A1}^{2} + 
                v_{1}^{2}\cos^{2}\theta(X-4) \right\} + \frac{1}{3}v_{A1}^{2}v_{1}^{2}\sin^{2}\theta X\\
                \times\left\{ \frac{v_{1}^{2}}{2}(5+X) + v_{A1}^{2}X(X-4) \right\} = 0,
\end{eqnarray}
and multiplying by $1/v_{A1}^{2}$ allows to express in terms of Alfv\'{e}n Mach number $M_{A1}$

\begin{eqnarray}\nonumber
        (v_{1}^{2}-Xv_{A1}^{2})^2 \left\{ 5\beta X + 2 M_{A1}^{2}\cos^{2}\theta(X-4) \right\}
                + v_{A1}^{2}v_{1}^{2}\sin^{2}\theta X\\
                \times\left\{ M_{A1}^{2}(5+X)+2X(X-4)\right\} = 0.
\end{eqnarray}
%
Finally, after some algebraic procedures one arrives to the following solution

\begin{eqnarray}\nonumber
        (M_{A1}^{2}-X)^{2}\left\{ 5\beta X + 2M_{A1}^{2}\cos\theta^{2}(X-4) \right\} + M_{A1}^{2}\sin\theta^{2}X\\
                \times\left\{ M_{A1}^{2}(5+X) + 2X(X-4) \right\} = 0,
\end{eqnarray}
which is equation~(7) presented in section~5.3. \cite{vrsnak2002b} also arrived to the same solution. We used
equation~(B12) to calculated the Alfv\'{e}n Mach number assuming two cases, horizontal ($\theta=0$) and
perpendicular ($\theta=90$) shock propagation.
For a horizontal case equation~(B12) is reduced to $M_{A}=\sqrt{X}$, while for a perpendicular situation to
$M_{A}=\sqrt{\frac{X(5+X+5\beta)}{2(4-X)}}$.



\end{document}